\documentclass{jmlr} 

\usepackage{longtable}

\usepackage{booktabs}
\usepackage[load-configurations=version-1]{siunitx} 

\usepackage{bm}
\usepackage{algorithm}
\usepackage{algorithmic}
\usepackage{multirow}

\theorembodyfont{\upshape}
\theoremheaderfont{\scshape}
\theorempostheader{:}
\theoremsep{\newline}

\jmlrworkshop{Preliminary work}

\title[Interpretable ICD Code Embeddings]{Interpretable ICD Code Embeddings with Self- and Mutual-Attention Mechanisms}

\author{\Name{Dixin Luo} \Email{dixin.luo@duke.edu}\\ 
      \addr Department of Electrical and Computer Engineering\\
      Duke University\\
      Durham, NC, USA 
      \AND
      \Name{Hongteng Xu} \Email{hongteng.xu@duke.edu}\\
      \addr Department of Electrical and Computer Engineering\\
      Duke University\\
      InfiniaML, Inc.\\
      Durham, NC, USA
      \AND
      \Name{Lawrence Carin} \Email{lcarin@duke.edu}\\
      \addr Department of Electrical and Computer Engineering\\
      Duke University\\
      Durham, NC, USA
      } 

\begin{document}

\maketitle

\begin{abstract}
We propose a novel and interpretable embedding method to represent the international statistical classification codes of diseases and related health problems ($i.e.$, ICD codes). 
This method considers a self-attention mechanism within the disease domain and a mutual-attention mechanism jointly between diseases and procedures. This framework captures the clinical relationships between the disease codes and procedures associated with hospital admissions, and it predicts procedures according to diagnosed diseases. 
A self-attention network is learned to fuse the embeddings of the diseases for each admission. 
The similarities between the fused disease embedding and the procedure embeddings indicate which procedure should potentially be recommended. 
Additionally, when learning the embeddings of the ICD codes, the optimal transport between the diseases and the procedures within each admission is calculated as a regularizer of the embeddings. 
The optimal transport provides a mutual-attention map between diseases and the procedures, which suppresses the ambiguity within their clinical relationships. 
The proposed method achieves clinically-interpretable embeddings of ICD codes, and outperforms state-of-the-art embedding methods in procedure recommendation.
\end{abstract}

\section{Introduction}
The International Classification of Diseases (ICD) is provided by the World Health Organization (WHO), and  
contains codes for diseases, procedures, and external causes of injury or disease. ICD codes play an important role in patient electrical health records (EHRs). 
For example, the hospital admission of a patient is often summarized as a set of disease ICD codes and a set of procedure ICD codes. 
The disease ICD codes represent the diagnosis provided by doctors, and the procedure ICD codes indicate the treatments applied to the patient. 

A significant and interesting problem is predicting procedures given diagnosed diseases, which can be useful for improving the effectiveness and the efficiency of hospital admission. 
From the viewpoint of machine learning, this problem can be formulated as an ICD code embedding task. 
{Specifically, given a patient admission record, we aim to represent the ICD codes of the diseases and procedures appearing in the record as embedding vectors.} 
Accordingly, one may anticipate a procedure for a given disease when the two have similar embedding vectors.

Unfortunately, most existing embedding methods may not be suitable for the proposed problem because of the special properties of ICD codes. 
For each admission, the corresponding disease ICD codes and procedure ICD codes are ranked according to a manually-defined priority, rather than their real clinical relationships. 
Additionally, a disease often leads to multiple procedures and a procedure may correspond to multiple diseases. 
In other words, the diseases and the procedures in an admission often have complicated clinical relationships, but the mapping between these two code sets is unavailable. 
Such uncertainties in the admission records are challenging for existing embedding methods, because they require well-structured observations, $e.g.$, sequential data like words in sentences~\citep{mikolov2013efficient,pennington2014glove}, pairwise interactions like user-item pairs in recommender systems~\citep{rendle2009bpr,chen2018sequential}, and node interactions in graphs~\citep{perozzi2014deepwalk,tang2015line,grover2016node2vec}. 
As a result, applying existing methods to embed ICD codes directly suffers from a high risk of model misspecification. 
Moreover, like other clinical data analysis tasks~\citep{choi2016multi,mullenbach2018explainable,mahmood2018automated}, the proposed embeddings of ICD codes should be interpretable, $e.g.$, it is desirable that the clinical relationships between diseases and procedures can be explicitly captured by the distance/similarity between their embedding vectors. 
Although some existing embedding methods can capture simple pairwise relationships between their embedded entities, it is still difficult to describe more complicated relationships among multiple entities.

\begin{figure}[t]
    \centering
    \includegraphics[width=0.9\linewidth]{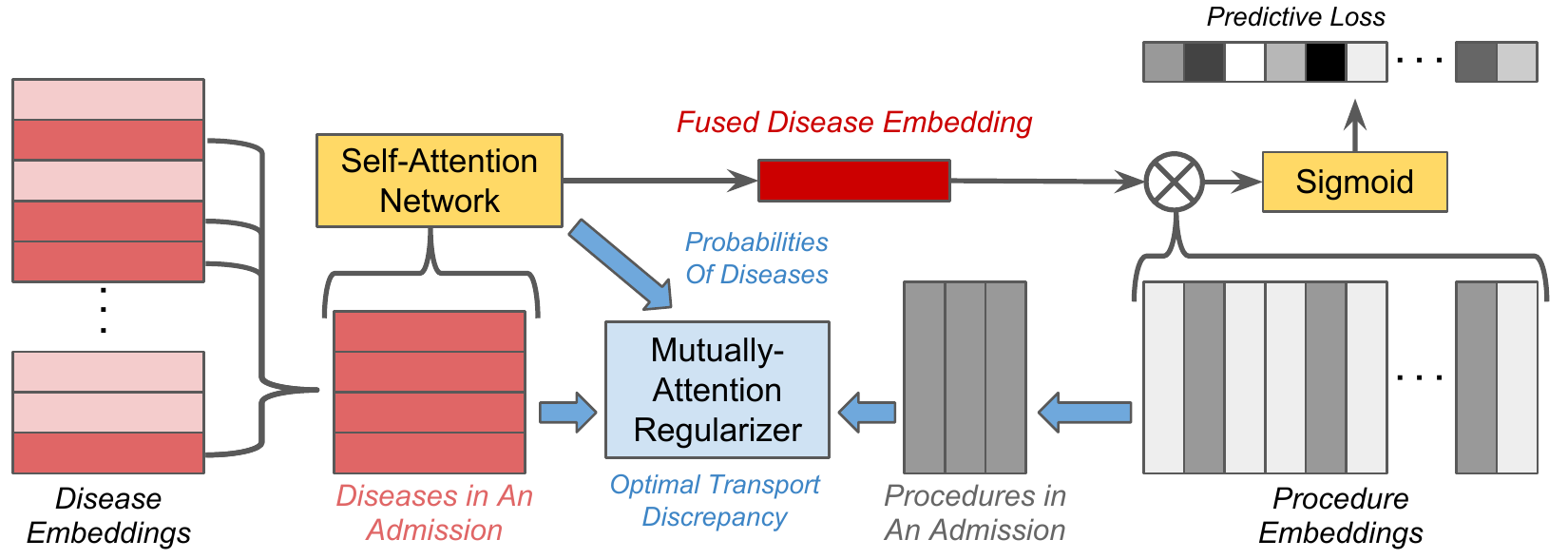}
    \caption{Scheme of proposed method. The mutual-attention module and the blue arrows are only used in the training phase. The remaining modules and the black arrows are used in both the training and testing phases.}
    \label{fig:scheme}
\end{figure}


Focusing on the challenges of ICD code embeddings, we propose an interpretable embedding method with novel self- and mutual-attention mechanisms. 
As illustrated in Figure~\ref{fig:scheme}, our method contains a self-attention network to fuse observed disease embeddings together. 
Accordingly, the similarity between the fused disease embedding and each procedure embedding is calculated, and procedures with high similarity are predicted to fit the observed procedures. 
When learning the embeddings and the self-attention network, we take advantage of the optimal-transport distance~\citep{villani2008optimal} between observed diseases and procedures as the regularizer of our model, achieving a mutual-attention map between them. 
The self- and the mutual-attention mechanisms are connected via the estimated probabilities of diseases. 
The proposed method has the advantages of interpretability of learned embeddings. 
Specifically, for each admission, the self-attention network estimates the probabilities of observed diseases, which can be interpreted as the significance of the disease in the admission.
Additionally, the mutual-attention regularizer estimates the optimal transport between the observed diseases and procedures, which estimates their clinical relationships. 

The proposed method can be used to recommend suitable procedures according to diagnosed diseases, which can be used to improve the efficiency of hospital admission, $i.e.$, for some typical diseases, clinicians, especially those junior and with limited clinical experience, can query suitable procedures quickly. 
Additionally, the recommended procedure codes can help to double-check the codes entered by clinicians.
For other clinical data analysis tasks like ICD code assignment~\citep{baumel2017multi,huang2018empirical}, which requires ICD code embeddings as the input of down-stream models and applications, the proposed embeddings generated by our method can either provide high-quality inputs or improve the training of their ICD code embeddings via good initialization.

\section{Target Data Set and Problem Statement}\label{sec:data}
The proposed work employs the publicly-available MIMIC-III dataset~\citep{johnson2016mimic}, developed by the MIT Lab for Computational Physiology.
It comprises over 58,000 hospital admissions recorded from June 2001 to October 2012 for 38,645 adults and 7,875 neonates. 
For each admission, its ICD codes are generated for billing purposes at the end of the hospital stay, which includes a set of disease ICD codes and a set of procedure ICD codes. 
The ICD codes employ the ICD-9 format.\footnote{\url{https://www.cdc.gov/nchs/icd/icd9.htm}}
In the MIMIC-III dataset, 14,567 disease ICD codes and 3,882 procedure ICD codes are observed. 
Each admission contains 1 to 41 diseases and 1 to 40 procedures. 

\begin{figure}[t]
    \centering
    \subfigure[Small set]{
    \includegraphics[width=0.31\linewidth]{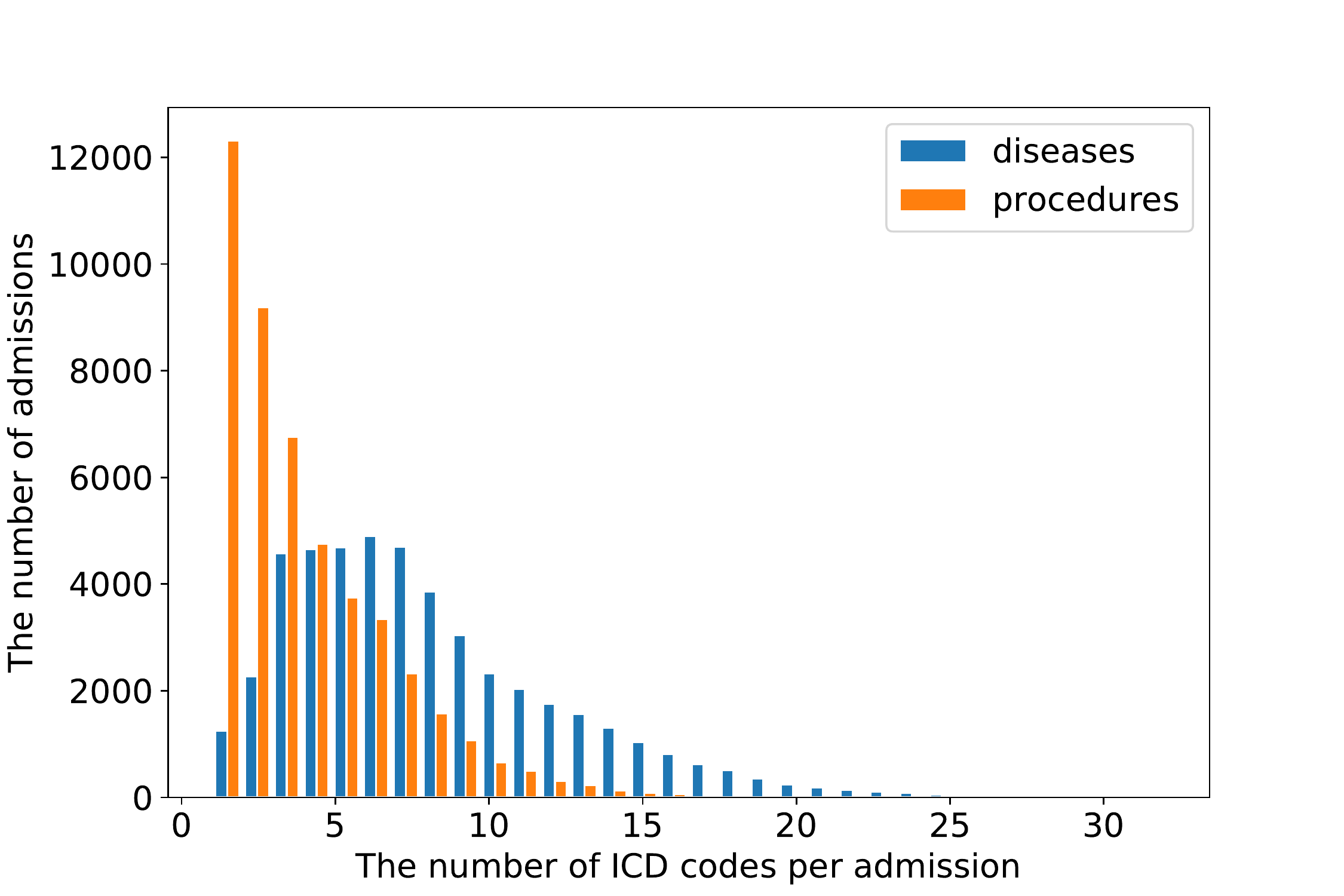}
    }
    \subfigure[Medium set]{
    \includegraphics[width=0.31\linewidth]{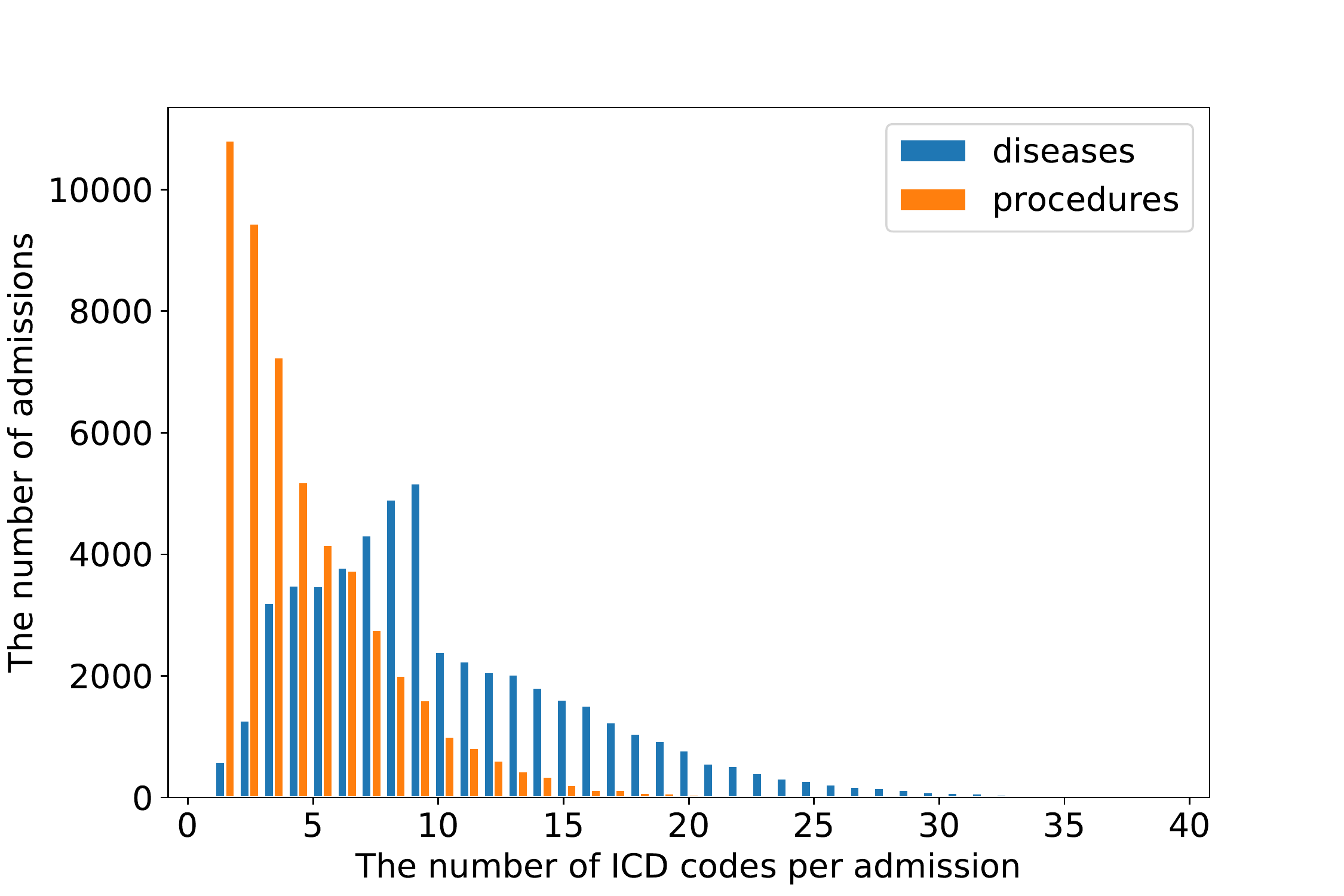}
    }
    \subfigure[Large set]{
    \includegraphics[width=0.31\linewidth]{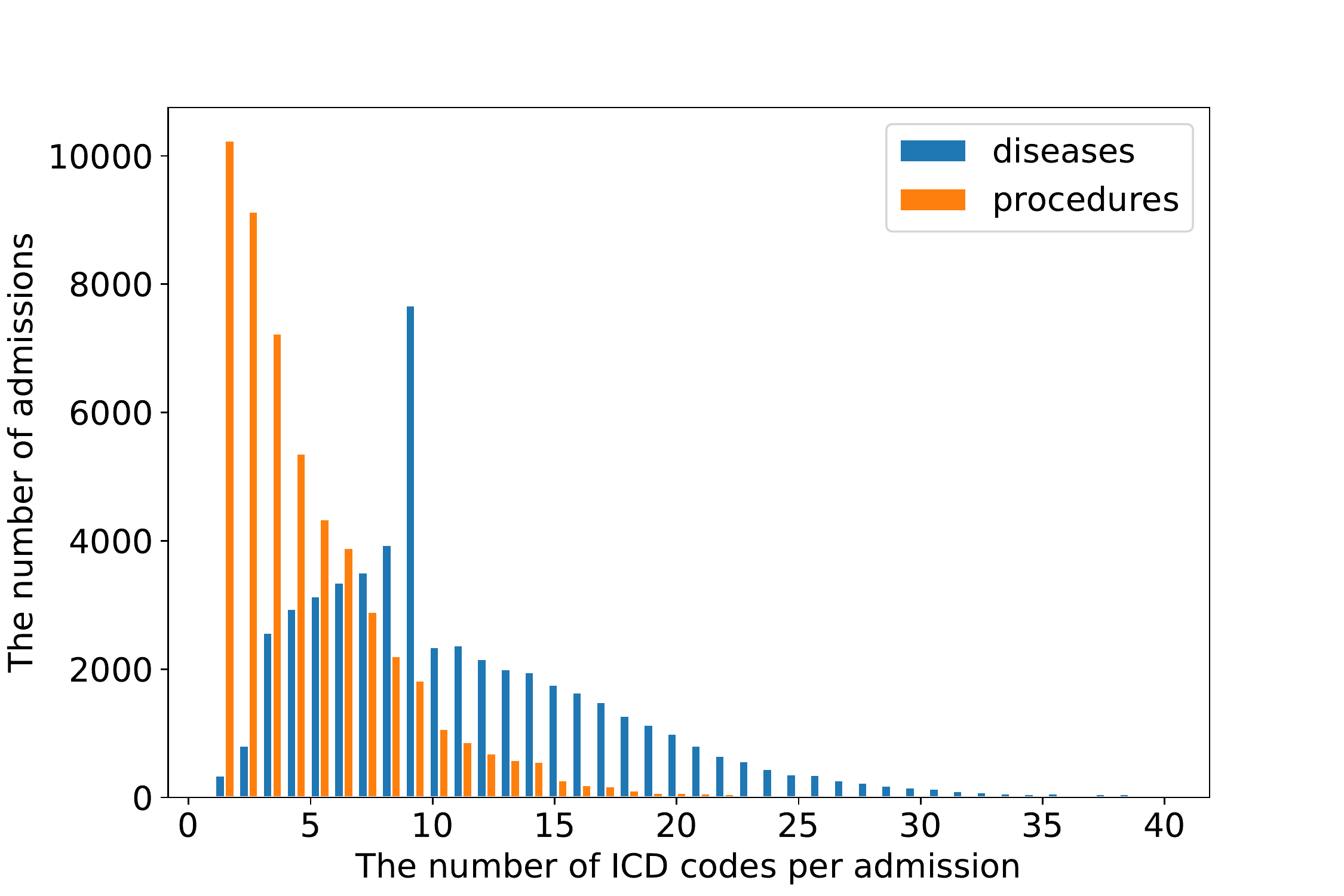}
    }
    \caption{Histograms of admissions, with respect to the number of diseases (blue) and procedures (orange) per admission for the three subsets.}
    \label{fig:histogram}
\end{figure}

We consider three subsets of the MIMIC-III dataset.
For each subset, we select the admissions having non-empty diseases and procedures. 
The small dataset contains 28,315 admissions with 247 diseases and 75 procedures, and each ICD code appears at least 500 times.
The medium dataset contains 30,555 admissions with 874 diseases and 258 procedures, and each ICD code appears at least 100 times.
The large dataset contains 31,213 admissions with 2,765 diseases and 819 procedures, and each ICD code appears at least 10 times.
For these three subsets, the histograms of admissions with respect to the number of diseases and that of procedures are shown in Figure~\ref{fig:histogram}.
We find that the histograms of admissions corresponding to the number of procedure ICD code per admission are consistent across different subsets, implying that the empirical distribution of procedures is stable and may yield an exponential distribution. 
On the other hand, the histograms of admissions with respect to the number of diseases per admission have changes with the increase of data size. 

We denote the set of disease and procedure ICD codes as $\mathcal{D}=\{d_1,d_2,...\}$ and $\mathcal{P}=\{p_1,p_2,...\}$, respectively, where the sizes of the two sets are $|\mathcal{D}|$ and $|\mathcal{P}|$.\footnote{Here, $|\cdot|$ represents the cardinality of the associated set, counting the number of elements.}
Each element $d\in\mathcal{D}$ represents a specific disease and each $p\in\mathcal{P}$ represents a specific procedure. 
Suppose that we observed $I$ hospital admissions.
For the $i$-th admission, $i\in\{1,...,I\}$, the diseases and the procedures associated with the admission are denoted $\mathcal{D}_i$ and $\mathcal{P}_i^+$, respectively. 
Here, $\mathcal{P}_i^+$ contains the ``positive'' procedures corresponding to $\mathcal{D}_i$. 
Accordingly, we can define/generate ``negative'' procedures that are potentially irrelevant to $\mathcal{D}_i$, $i.e.$, $\mathcal{P}_i^-\subset \mathcal{P}\setminus\mathcal{P}_i^+$, used in the following learning algorithm. 

Given such observations, we aim to {$i$) embed ICD codes of diseases and procedures; $ii$) predict reasonable procedures from diagnosed diseases according to proposed embeddings; and $iii$) it is desired that the prediction is clinically-interpretable}.
In the following sections, we denote the embeddings of diseases and procedures as $\bm{U}=[\bm{u}_d]\in \mathbb{R}^{M\times |\mathcal{D}|}$ and $\bm{V}=[\bm{v}_p]\in\mathbb{R}^{M\times |\mathcal{P}|}$, respectively, where $M$ is the embedding dimension.

\section{Proposed Model and Learning Algorithm}

\subsection{Predicting procedures from diseases}
Our embedding method learns a predictive model of procedures. 
For the $i$-th admission, we predict the set of procedures $\mathcal{P}_i^+$ from the set of diseases $\mathcal{D}_i$. 
Specifically, for each procedure $p\in\mathcal{P}$, we represent the probability of $p$ conditioned on $\mathcal{D}_i$ via the following parametric model:
\begin{eqnarray}\label{eq:prob}
\begin{aligned}
\mbox{Prob}_{\theta}(p | \mathcal{D}_i)=\mbox{sigmoid}(\bm{v}_p^{\top}f(\bm{U}_{\mathcal{D}_i})),~\forall~p\in\mathcal{P}.
\end{aligned}
\end{eqnarray}
where $\bm{v}_p\in\mathbb{R}^{M}$ is the embedding of the procedure $p$, $\bm{U}_{\mathcal{D}_i}=[\bm{u}_d]_{d\in\mathcal{D}_i}\in\mathbb{R}^{M\times |\mathcal{D}_i|}$ contains the columns of $\bm{U}$ corresponding to $\mathcal{D}_i$, and $f(\cdot)$ is a function fusing columns of $\bm{U}_{\mathcal{D}_i}$ to a single vector. 
For convenience, we denote the parameters of the predictive model as $\theta=\{\bm{U}\in \mathbb{R}^{M\times |\mathcal{D}|},~\bm{V}\in\mathbb{R}^{M\times |\mathcal{P}|},~f(\cdot)\}$.

For the positive procedure $p\in \mathcal{P}_i^+$, it is desirable that the proposed model {makes} $\mbox{Prob}_{\theta}(p | \mathcal{D}_i)$ approach $1$. 
By contrast, for the negative procedure $p\in\mathcal{P}\setminus\mathcal{P}_i^+$, the proposed model should suppress $\mbox{Prob}_{\theta}(p | \mathcal{D}_i)$ to $0$. 
It should be noted that for each admission the number of negative procedures is much larger than that of positive procedures ($i.e.$, $|\mathcal{P}\setminus\mathcal{P}_i^+|\gg |\mathcal{P}_i^+|$).
Therefore, in practice we must apply the negative sampling strategy used in other embedding methods~\citep{rendle2009bpr,mikolov2013efficient}, randomly selecting a subset of negative procedures, $i.e.$, $\mathcal{P}_i^-\subset \mathcal{P}\setminus\mathcal{P}_i^+$ and ensuring $|\mathcal{P}_i^-|=|\mathcal{P}_i^+|$.

Accordingly, we can learn the model via maximum likelihood estimation (MLE): for the $i$-th admission, its predictive loss is 
\begin{eqnarray}\label{eq:loss_p}
\begin{aligned}
L_{P}(\mathcal{D}_i,\mathcal{P}_{i}^{+},\mathcal{P}_{i}^{-};\theta)=-\sideset{}{_{p\in \mathcal{P}_i^{+}}}\sum \log \mbox{Prob}_{\theta}(p|\mathcal{D}_i) - \sideset{}{_{p\in\mathcal{P}_i^{-}}}\sum\log (1-\mbox{Prob}_{\theta}(p|\mathcal{D}_i)).
\end{aligned}
\end{eqnarray}
By minimizing $L_P$, we maximize the log-likelihood of the observed procedures given the corresponding diagnosed diseases, and suppress the log-likelihood of the irrelevant procedures.

\subsection{Fusing disease embeddings with a self-attention mechanism}
The key of the above predictive model is the fusion function $f(\cdot)$, which has a large influence on the interpretability of the model and the final performance on prediction.
The simplest fusion strategy is average pooling, $i.e.$, $f(\bm{U}_{\mathcal{D}_i})=\frac{1}{|\mathcal{D}_i|}\sum_{d\in\mathcal{D}_i}\bm{u}_d$, which (questionably) assumes that the different diseases have the same influence on the prediction of procedures. 
Another strategy is max-pooling, which keeps the maximum value for each feature dimension while ignoring the contributions from other diseases. 
To overcome the problems of the two strategies above and improve the performance of our predictive model, we propose a novel self-attention network as the fusion function, achieving an adaptive fusion strategy for disease embeddings. 

\begin{figure}[t]
    \centering
    \subfigure[Self-attention]{
    \includegraphics[height=4.7cm]{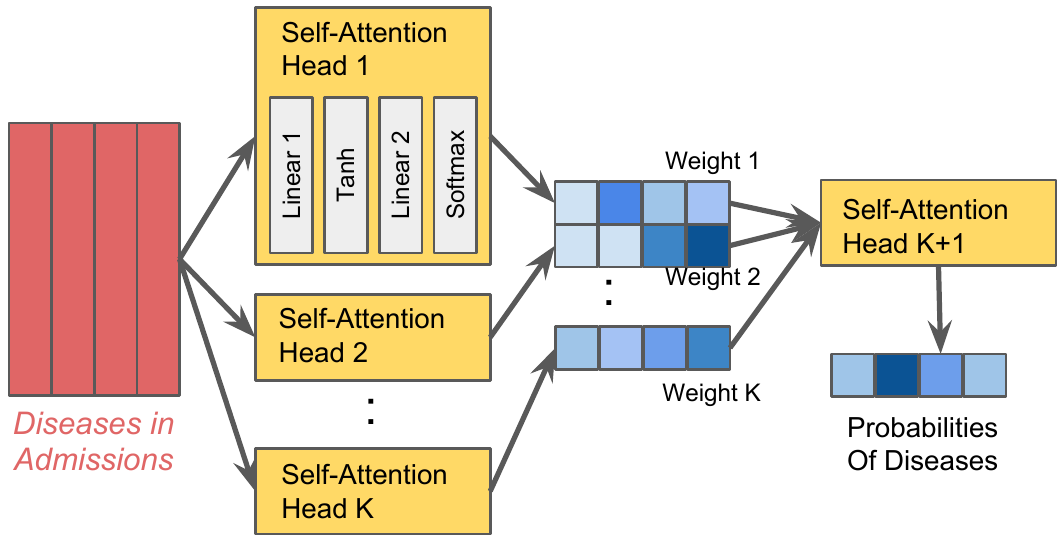}\label{fig:self_att}
    }
    \subfigure[mutual-attention]{
    \includegraphics[height=4.7cm]{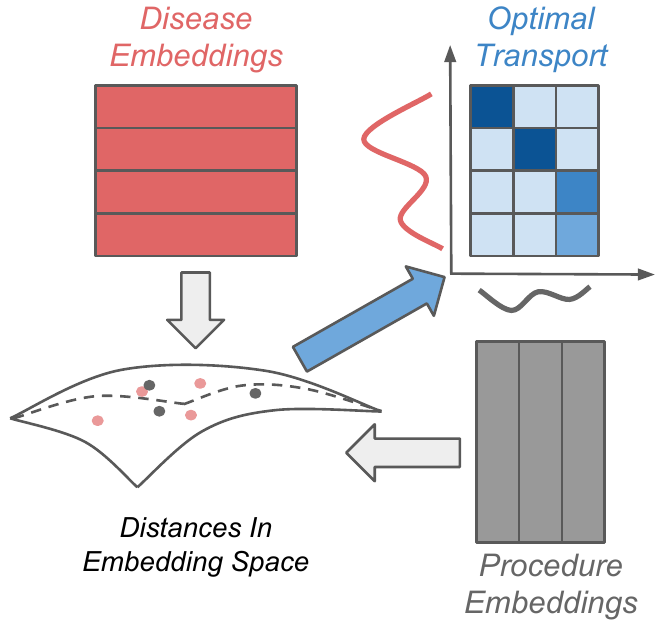}\label{fig:mutual_att}
    }
    \caption{Illustration of the self- and mutual-attention mechanisms in our model. In (a), each self-attention head is a feed-forward neural network, whose architecture is shown in head 1.}
    \label{fig:modules}
\end{figure}

The proposed self-attention network is inspired by the multi-head attention architecture in~\citep{vaswani2017attention} and the self-attentive embedding structure in~\citep{lin2017structured}. 
As shown in Figure~\ref{fig:self_att}, our self-attention network is a two-layer architecture.
The first layer contains $K$ heads, each of which is a self-attention function that generates a weight vector $\bm{w}_k\in\Sigma^{|\mathcal{D}_i|}$ from $\bm{U}_{\mathcal{D}_i}$:\footnote{$\Sigma^{N}$ represents the $N$-dimensional simplex.}
\begin{eqnarray}\label{eq:sa1}
\begin{aligned}
\bm{w}_k = \mbox{softmax}(\bm{a}_k^{\top}\mbox{tanh}(\bm{A}_k\bm{U}_{\mathcal{D}_i})),~\forall~k=1, ..., K,
\end{aligned}
\end{eqnarray}
where $\bm{A}_k\in\mathbb{R}^{M\times M}$ and $\bm{a}_k\in\mathbb{R}^{M}$ are the parameters of the $k$-th head.
The second layer is a single self-attention function, which takes the concatenation of $\bm{w}_k$ as input and generates the final weight vector.
Denote $\bm{W}=\mbox{concatenate}(\bm{w}_1,...,\bm{w}_K)\in\mathbb{R}^{K\times |\mathcal{D}_i|}$, the final weight is derived via 
\begin{eqnarray}\label{eq:sa2}
\begin{aligned}
\bm{\mu}_{\mathcal{D}_i} = \mbox{softmax}(\bm{b}^{\top}\mbox{tanh}(\bm{B}\bm{W})),
\end{aligned}
\end{eqnarray}
where $\bm{B}\in\mathbb{R}^{K\times K}$ and $\bm{b}\in\mathbb{R}^{K}$ is the parameter of this layer.
For convenience, we represent the whole process as $\bm{\mu}_{\mathcal{D}_i}=\mbox{MHA}(\bm{U}_{\mathcal{D}_i})\in\Sigma^{|\mathcal{D}_i|}$, and accordingly $f(\bm{U}_{\mathcal{D}_i})=\bm{U}_{\mathcal{D}_i}\bm{\mu}_{\mathcal{D}_i}$. 

From a probabilistic viewpoint, $\bm{\mu}_{\mathcal{D}_i}$ may be interpreted as a distribution of the diseases in $\mathcal{D}_i$: the higher probability a disease has, the more significant influence the disease has on predicting procedures. 

\subsection{Learning with an optimal transport-based mutual-attention mechanism}
Besides the predictive loss in (\ref{eq:loss_p}), we further design a regularizer based on the optimal-transport distance between observed diseases and procedures, achieving a mutual-attention mechanism in the training phase. 
As mentioned above, a disease may lead to multiple procedures and a procedure can be shared by different diseases. 
Therefore, for the $i$-th admission, there exists a complicated map from $\mathcal{D}_i$ to $\mathcal{P}_i^+$. 
We estimate this map explicitly via minimizing the optimal-transport distance~\citep{villani2008optimal} between $\mathcal{D}_i$ and $\mathcal{P}_i^+$, and the estimated map is used to regularize the learning of embeddings.

Denote the distributions of diseases and corresponding procedures in the $i$-th admission as $\bm{\mu}_{\mathcal{D}_i}$ and $\bm{\mu}_{\mathcal{P}_i^+}$, respectively. 
Here, $\bm{\mu}_{\mathcal{D}_i}$ is the weighted vector derived via our self-attention networks, and the $\bm{\mu}_{\mathcal{P}_i^+}$ is assumed to be a uniform distribution. 
 {As illustrated in Figure~\ref{fig:mutual_att},} given the embeddings of the diseases and the procedures, $i.e.$, $\bm{U}_{\mathcal{D}_i}$ and $\bm{V}_{\mathcal{P}_i^+}$, we calculate the distance matrix between them, denoted $\bm{C}_{\theta}(\mathcal{D}_i, \mathcal{P}_i^{+}) = [c_{dp}]\in\mathbb{R}^{|\mathcal{D}_i|\times |\mathcal{P}_i^+|}$. 
Because the probability in (\ref{eq:prob}) is calculated based on the inner product between disease embedding and procedure embedding, we calculate the elements of the distance matrix based on the cosine similarity (normalized inner product) between the embeddings:
\begin{eqnarray}\label{eq:cost}
\begin{aligned}
c_{dp} = 1 - \frac{\bm{u}_d^{\top}\bm{v}_{p}}{\|\bm{u}_d\|_2\|\bm{v}_p\|_2},~\forall~d\in\mathcal{D}_i~\mbox{and}~p\in\mathcal{P}_i^+.
\end{aligned}
\end{eqnarray}

Based on the distance matrix, the optimal-transport distance provides a measure of the dissimilarity between $\bm{\mu}_{\mathcal{D}_i}$ and $\bm{\mu}_{\mathcal{P}_i^+}$, defined as
\begin{eqnarray}\label{eq:loss_ot}
\begin{aligned}
L_{OT}(\mathcal{D}_i,\mathcal{P}_{i}^{+};\theta)=\sideset{}{_{\bm{T}\in \Pi(\bm{\mu}_{\mathcal{D}_i},\bm{\mu}_{\mathcal{P}_i^{+}})}}\min\langle \bm{C}_{\theta}(\mathcal{D}_i, \mathcal{P}_i^{+}), \bm{T}\rangle,
\end{aligned}
\end{eqnarray}
where $\Pi(\bm{\mu}_{\mathcal{D}_i},\bm{\mu}_{\mathcal{P}_i^{+}})=\{\bm{T}|\bm{T}\bm{1}_{|\mathcal{P}_i^{+}|}=\bm{\mu}_{\mathcal{D}_i},~\bm{T}^{\top}\bm{1}_{|\mathcal{D}_i|}=\bm{\mu}_{\mathcal{P}_i^{+}}\}$ is the set of all possible joint distributions with $\bm{\mu}_{\mathcal{D}_i}$ and $\bm{\mu}_{\mathcal{P}_i^+}$ as marginals;
$\langle\cdot,\cdot\rangle$ represents the inner product between matrices.
The optimal transport, $i.e.$, $\widehat{\bm{T}}_i=\arg\min_{\bm{T}\in \Pi(\bm{\mu}_{\mathcal{D}_i},\bm{\mu}_{\mathcal{P}_i^{+}})}\langle \bm{C}_{\theta}(\mathcal{D}_i, \mathcal{P}_i^{+}), \bm{T}\rangle$, is the joint distribution that minimizes the distance between $\bm{\mu}_{\mathcal{D}_i}$ and $\bm{\mu}_{\mathcal{P}_i^+}$. 
A detailed introduction to the optimal-transport distance is found in Appendix~\ref{ap:1} and within~\citep{villani2008optimal,cuturi2013sinkhorn,benamou2015iterative}.

With the help of the optimal-transport distance, we achieve a mutual-attention mechanism for embedding ICD codes: the optimal transport $\widehat{\bm{T}}_i$ explicitly represents the clinical relationships between $\mathcal{D}_i$ and $\mathcal{P}_i^+$ -- if disease $d\in\mathcal{D}_i$ mainly yields procedure $p\in\mathcal{P}_i^+$, element $t_{dp}$ in $\widehat{\bm{T}}$ will ideally have a corresponding large value.

\subsection{Proposed learning method}
Given a set of admissions, $\{(\mathcal{D}_i, \mathcal{P}_i^{+}, \mathcal{P}_i^{-})\}_{i=1}^{I}$, we jointly consider the predictive loss in (\ref{eq:loss_p}) and the optimal-transport distance in (\ref{eq:loss_ot}), and learn the proposed model by solving the following optimization problem:
\begin{eqnarray}\label{eq:opt}
\begin{aligned}
\sideset{}{_{\theta}}\min \sideset{}{_{i=1}^{I}}\sum L_{P}(\mathcal{D}_i,\mathcal{P}_{i}^{+},\mathcal{P}_{i}^{-};\theta) + \alpha L_{OT}(\mathcal{D}_i,\mathcal{P}_{i}^{+};\theta) {.}
\end{aligned}
\end{eqnarray}
This is a nested minimization problem, because the second term involves the optimization of optimal transport. 
We solve this problem via alternating optimization, which involves the following two steps:
\begin{itemize}
    \item\textbf{Update optimal transport.} For each admission, given current embeddings $\bm{U}_{\mathcal{D}_i}$ and $\bm{V}_{\mathcal{P}_i^+}$, and the estimated distribution of $\bm{\mu}_{\mathcal{D}_i}$, we update the optimal transport by solving (\ref{eq:loss_ot}), $i.e.$, $\widehat{\bm{T}}_i=\arg\sideset{}{_{\bm{T}\in \Pi(\bm{\mu}_{\mathcal{D}_i},\bm{\mu}_{\mathcal{P}_i^{+}})}}\min\langle \bm{C}_{\theta}(\mathcal{D}_i, \mathcal{P}_i^{+}), \bm{T}\rangle$.
    This optimization problem can be solved effectively via the proximal gradient method~\citep{xie2018fast}, with linear convergence. 
    The detailed introduction of the proximal gradient method is given in Appendix~\ref{ap:1}.
    \item\textbf{Update embeddings and self-attention network.} Given updated optimal transport matrices $\{\widehat{\bm{T}}_i\}_{i=1}^{I}$, we plug them into the second term of (\ref{eq:opt}), and update the ICD code embeddings and the self-attention network via mini-batch gradient descent. 
    Specifically, we solve the following problem:
    \begin{eqnarray}\label{eq:update_emb}
    \begin{aligned}
    \sideset{}{_{\theta}}\min \sideset{}{_{i\in \mathcal{B}}}\sum L_{P}(\mathcal{D}_i,\mathcal{P}_{i}^{+},\mathcal{P}_{i}^{-};\theta) + \alpha \langle \bm{C}_{\theta}(\mathcal{D}_i, \mathcal{P}_i^{+}), \widehat{\bm{T}}_i\rangle {,}
    \end{aligned}
    \end{eqnarray}
    where $\mathcal{B}$ represents a batch of admissions. 
\end{itemize}
The learning algorithm is summarized in Algorithm~\ref{alg1}. 
Note that the updating of optimal transport for different admissions can be done in parallel, and the gradient of (\ref{eq:update_emb}) can be calculated efficiently via backpropagation on graphical processing units (GPUs). 

\begin{algorithm}[t]
	\caption{Proposed ICD code embedding method}
	\label{alg1}
	\begin{algorithmic}[1]
		\STATE \textbf{Input:} Observed admissions $\{\mathcal{D}_i,\mathcal{P}_i^+\}_{i=1}^{I}$. The dimension of embeddings $M$. The number of self-attention heads $K$. The weight of regularizer $\alpha$.
		\STATE \textbf{Output:} The parameters of proposed model $\theta$.
		\STATE Initialize $\theta=\{\bm{U},\bm{V},f(\cdot)\}$ randomly.
        \WHILE{not converge}
        \STATE Sample a batch of admissions randomly, $i.e.$, $\{\mathcal{D}_i,\mathcal{P}_i^+\}_{i\in\mathcal{B}}$.
        \STATE Sample negative procedures $\mathcal{P}_i^-$ for each admission, and get $\{\mathcal{D}_i,\mathcal{P}_i^+,\mathcal{P}_i^-\}_{i\in\mathcal{B}}$. 
        \STATE \CommentSty{Update optimal transport matrices:}
        \FOR{$i\in \mathcal{B}$}
        \STATE Solve (\ref{eq:loss_ot}) via proximal gradient method.
        \ENDFOR
        \STATE \CommentSty{Update model's parameters:}
        \STATE Solve (\ref{eq:update_emb}) via mini-batch gradient descent.
        \ENDWHILE
	\end{algorithmic}
\end{algorithm}

\section{Related Work}
\subsection{Embedding techniques}
Embeddings have been widely used to represent and analyze real-world entities. 
For user-item pairs in recommendation systems, low-rank factorization models are applied to estimate user and item embeddings from their observed interactions~\citep{herlocker1999algorithmic,rendle2009bpr}. 
When side information is available, such as reviews of items provided by users or the images of items, the embeddings can be further parametrized via neural networks~\citep{kang2018self,chen2018sequential,chenxu2018adversarial}. 
For natural language processing, word embeddings have been widely used to represent words in sentences. 
Typical word embedding techniques include Word2Vec \citep{mikolov2013efficient} and GloVe \citep{pennington2014glove}, which maximize the coherency of the words and their contexts in sentences. 
Following the same strategy, many node embedding methods have been proposed to represent nodes in graphs, $e.g.$, DeepWalk \citep{perozzi2014deepwalk}, LINE \citep{tang2015line}, and Node2Vec \citep{grover2016node2vec}. 
Most of these embedding methods employ the same framework -- first generate sequential observations through random walks in the graph, and then apply word embeddings by maximizing the coherency of adjacent nodes on the random walks. 
Focusing on clinical data analysis, embedding techniques have been applied to many tasks, like ICD code assignment~\citep{shi2017towards,baumel2017multi,mullenbach2018explainable,huang2018empirical}, clinical data clustering, and treatment prediction~\citep{bajor2018embedding,harutyunyan2017multitask,choi2016multi}. 
Recently, the work in~\citep{xu2018distilled,xu2019gromov} makes efforts to learn ICD code embeddings directly through admission records based on optimal transport-based methods. 
However, the methods don't scale well, and therefore it is hard to apply in practical systems.

\subsection{Attention models}
Attention models have proven useful for many machine learning tasks. 
Such models are typically applied to a set or sequence of vectors, and the associated weights on the vectors characterize their importance for a given task~\citep{vinyals2015pointer}. 
The weights highlight the important parts of the sequence of vectors, and effectively provide an adaptive pooling strategy to obtain a global representation of all observations~\citep{lin2017structured}. 
A successful example of attention models is its application to natural language processing (NLP) tasks, like question-answering and document comprehension~\citep{devlin2018bert,choi2017coarse} -- the attention model has been core to advanced NLP modules like the  ``transformer''~\citep{vaswani2017attention}.
Leading language models like GPT~\citep{radford2018improving} and BERT~\citep{devlin2018bert} also rely on various attention mechanisms. 
Besides NLP, attention models have recently been applied to other tasks, including recommendation systems~\citep{kang2018self}, imitation learning~\citep{kool2018attention}, and multi-instance learning~\citep{ilse2018attention}. 
Attention models are also applicable for healthcare, a, $e.g.$, ECG rhythm classification~\citep{goodfellow2018towards}, admission prediction~\citep{choi2016retain}, and heart failure prediction~\citep{choi2017gram}.

\subsection{Optimal transport-based learning}
Learning based on the optimal-transport distance has recently attracted much attention, such as in distribution estimation~\citep{boissard2015distribution} and clustering~\citep{ye2017fast}.
It can also be used as a loss function when learning deep generative models~\citep{courty2017learning,arjovsky2017wasserstein}.
The main bottleneck of the application of optimal-transport distance is its high computational complexity. 
This problem has been eased greatly by the Sinkhorn-Knopp algorithm \citep{sinkhorn1967concerning}. 
Specifically, by adding an entropy regularizer~\citep{cuturi2013sinkhorn}, the optimal-transport distance can be approximated via iterative Bregman projection~\citep{benamou2015iterative}. 
The algorithm achieves near-linear-time complexity~\citep{altschuler2017near}, and its convergence and stability can be further improved via the inexact proximal point method~\citep{xie2018fast}.
However, the application of optimal transport to problems in healthcare has not been widely investigated.

\section{Experiments} 
\subsection{Comparisons on procedure recommendation}
To demonstrate the effectiveness of the proposed embedding method, we test it on the three datasets introduced in Section~\ref{sec:data} and compare it with its variants and existing embedding methods.   
Specifically, we denote our \textbf{E}mbedding method with \textbf{S}elf- and \textbf{M}utual-\textbf{A}ttention mechanisms as \textbf{E+SA+MA}. 
Its variants include: 1) setting the fusion function to a max-pooling function, and learning without optimal transport-based mutual-attention regularizer ($i.e.$, \textbf{E+Pooling}); 2) using the proposed self-attention network as the fusion function, but learning without the mutual-attention regularizer ($i.e.$, \textbf{E+SA}); and 3) using max-pooling as the fusion function but considering the mutual-attention regularizer ($i.e.$, \textbf{E+MA}). 
These variants provide understanding of the significance of the self- and mutual-attention mechanisms. 
The key hyperparameters of the proposed method include the dimension of the embedding vector $M$, the weight of the mutual-attention regularizer $\alpha$ and the number of heads in the self-attention network $K$. 
In all models we set $M=200$, $\alpha=0.1$, and $K=8$. 
The robustness of our method to these hyperparameters is analyzed in Section~\ref{ssec:robust}. 
All the methods learn ICD code embeddings via Adam~\citep{kingma2014adam}, setting the learning rate to $0.001$, batch size as $300$, and the number of epochs as $25$.

We also consider the following methods as baselines: 1) the \textbf{Word2Vec} method~\citep{mikolov2013efficient}, which enumerates all possible disease-procedure pairs in each admission, and learns the ICD code embeddings to maximize the log-likelihood of the pairs; 2) the classical Bayesian personalized ranking method (\textbf{BPR})~\citep{rendle2009bpr}, which maximizes the coherency between the averaged disease embedding and the procedure embeddings in each admission; 3) the distilled Wasserstein method (\textbf{DWL})~\citep{xu2018distilled}, that learns the ICD code embeddings via a hierarchical optimal transport-based method; and 4) the Gromov-Wasserstein learning method (\textbf{GWL})~\citep{xu2019gromov}, that learns the ICD code embeddings with structural regularization on the embeddings within the domain of diseases and that of procedures.
Similar to our method, these baselines also learn ICD code embeddings with dimension $M=200$.

\begin{table}[t]
	\centering
	\caption{Comparison for various methods on recommendation results}\label{tab:cmp}
	\begin{small}
	\begin{tabular}{@{\hspace{1pt}}c@{\hspace{2pt}}|@{\hspace{2pt}}c@{\hspace{2pt}}
	|c@{\hspace{4pt}}c@{\hspace{4pt}}c
	|c@{\hspace{4pt}}c@{\hspace{4pt}}c
	|c@{\hspace{4pt}}c@{\hspace{4pt}}c
	|c@{\hspace{4pt}}c@{\hspace{4pt}}c@{\hspace{1pt}}}
	\hline\hline
        \multirow{2}{*}{Dataset} 
	    &\multirow{2}{*}{Method} 
	    &\multicolumn{3}{c|}{Top-1 (\%)} 
	    &\multicolumn{3}{c|}{Top-3 (\%)} 
	    &\multicolumn{3}{c|}{Top-5 (\%)} 
	    &\multicolumn{3}{c}{Top-10 (\%)}\\
	    \cline{3-14}
		& &R &P &F1 &R &P &F1 &R &P &F1 &R &P &F1\\ 
		\hline
	               &Word2Vec & 19.5 & 47.8 & 24.7
	                         & 35.4 & 34.9 & 30.8 
	                         & 47.1 & 29.6 & 32.0
	                         & 62.3 & 21.1 & 28.5\\
	    28,315     &DWL      & 19.7 & 48.2 & 25.0
	                         & 35.9 & 35.2 & 31.3
	                         & 47.5 & 30.3 & 32.4
	                         & 63.0 & 20.9 & 28.7\\
		Admissions &GWL      & 13.1 & 44.9 & 18.0
		                     & 19.2 & 25.2 & 18.8 
		                     & 23.9 & 21.6 & 19.5
		                     & 40.1 & 20.0 & 24.3\\
		247        &BPR      & 23.5 & 57.6 & 29.8
		                     & 44.8 & 43.5 & 38.7
		                     & 56.8 & 35.7 & 38.8 
		                     & 73.1 & 24.8 & 33.6\\
		Diseases   &E+Pooling& 24.5 & 58.5 & 30.9 
		                     & 46.2 & 44.0 & 39.6 
		                     & 58.0 & 35.7 & 39.1 
		                     & 74.5 & 25.1 & 34.0\\
		75         &E+SA     & 24.4 & 58.4 & 30.8 
		                     & 45.8 & 43.7 & 39.2 
		                     & 57.2 & 35.7 & 38.9 
		                     & 73.6 & 24.9 & 33.7\\
		Procedures &E+MA     & 23.7 & 57.4 & 30.0 
		                     & 45.2 & 43.2 & 38.8 
		                     & 57.4 & 35.6 & 38.8 
		                     & 74.2 & 25.0 & 33.8\\
		           &E+SA+MA  &\textbf{24.8} &\textbf{59.7} &\textbf{31.3} 
		                     &\textbf{46.4} &\textbf{44.1} &\textbf{39.7} 
		                     &\textbf{58.7} &\textbf{36.3} &\textbf{39.7} 
		                     &\textbf{74.9} &\textbf{25.3} &\textbf{34.2}\\ 
		           \hline
		           &Word2Vec & 7.8  & 27.6 & 10.9
		                     & 27.7 & 30.5 & 25.1
		                     & 38.3 & 26.9 & 27.7
		                     & 52.8 & 20.1 & 26.1\\
	    30,535     &DWL      & 8.0  & 27.5 & 11.1
	                         & 27.9 & 30.8 & 25.2
	                         & 39.5 & 27.0 & 27.9
	                         & 53.9 & 20.9 & 27.4\\
		Admissions &GWL      & 9.2  & 36.9 & 13.2
		                     & 11.8 & 15.9 & 11.6
		                     & 12.7 & 10.9 & 10.0
		                     & 15.8 & 7.8  & 8.9\\
		874        &BPR      & 10.2 & 35.8 & 14.9
		                     & 38.6 & 40.2 & 34.3 
		                     & 49.3 & 33.3 & 34.9
		                     & 65.2 & 23.8 & 31.4\\
		Diseases   &E+Pooling& 10.1 & 35.4 & 14.3 
		                     & 38.0 & 39.7 & 33.9 
		                     & 50.0 & 33.6 & 35.3 
		                     & 65.6 & 24.1 & 31.7\\
		258        &E+SA     & 18.2 & 50.0 & 23.6 
		                     & 36.6 & 39.1 & 33.0 
		                     & 48.7 & 33.1 & 34.6 
		                     & 66.0 & 24.1 & 31.8\\
		Procedures &E+MA     & 12.4 & 32.4 & 15.7 
		                     & 38.7 & 39.3 & 34.0 
		                     & 50.9 & 33.9 & 35.7 
		                     & 67.6 & 24.7 & 32.5\\
		           &E+SA+MA  &\textbf{20.3} &\textbf{53.1} &\textbf{26.1} 
		                     &\textbf{40.7} &\textbf{42.3} &\textbf{36.2} 
		                     &\textbf{53.0} &\textbf{35.4} &\textbf{37.2} 
		                     &\textbf{68.9} &\textbf{25.1} &\textbf{33.1}\\
		           \hline
	               &Word2Vec & 5.3  & 22.9 & 8.7 
	                         & 14.6 & 21.1 & 15.3 
	                         & 24.8 & 21.0 & 20.1 
	                         & 41.1 & 17.7 & 22.2\\
	    31,213     &DWL      & 5.6  & 23.0 & 9.0 
	                         & 14.9 & 21.3 & 15.6
	                         & 24.8 & 21.4 & 20.5
	                         & 42.0 & 18.2 & 23.0\\
		Admissions &GWL      & 5.5  & 19.0 & 7.5
		                     & 5.6  & 6.6  & 5.1
		                     & 5.7  & 4.2  & 4.1
		                     & 6.0  & 2.3  & 2.9\\
		2,765      &BPR      & 7.3  & 26.7 & 10.2
		                     & 23.0 & 27.1 & 21.2 
		                     & 38.4 & 27.6 & 27.9
		                     & 56.6 & 21.7 & 28.0\\
		Diseases   &E+Pooling& 7.4  & 27.3 & 10.4 
		                     & 16.5 & 23.3 & 17.1 
		                     & 38.7 & 27.3 & 27.8
		                     & 58.3 & 22.0 & 28.6\\
		819        &E+SA     & 8.0  & 28.2 & 11.1 
		                     & 20.0 & 26.0 & 19.7 
		                     & 36.5 & 26.5 & 26.6 
		                     & 56.6 & 21.7 & 28.0\\
		Procedures &E+MA     & 6.9  & 25.7 & 9.7 
		                     & 18.5 & 22.7 & 17.5 
		                     & 33.2 & 23.4 & 23.8 
		                     & 56.5 & 21.5 & 27.8\\
		           &E+SA+MA  &\textbf{8.5}  &\textbf{27.9} &\textbf{11.4} 
		                     &\textbf{23.1} &\textbf{27.2} &\textbf{21.4} 
		                     &\textbf{39.1} &\textbf{27.9} &\textbf{28.4} 
		                     &\textbf{60.0} &\textbf{22.8} &\textbf{29.6}\\
	\hline\hline
	\end{tabular}
    \end{small}
\end{table}

For each dataset, we use 80\% of the admissions for training and 20\% for testing. 
In the testing phase, we evaluate each embedding method for the task of recommending procedures according to diagnosed diseases. 
Similar to other works of recommendation systems, we evaluate the performance of each method in the testing phase via top-$L$ precision, recall, and F1-score. 
For the $i$-th testing admission, $i\in\{1,...,I\}$, each method recommends a set of procedures with length $L$, denoted $\widehat{\mathcal{P}}_i^{+}$. 
Given the ground-truth set $\mathcal{P}_i^+$, we calculate the top-$L$ precision, recall and F1-score as follows:
\begin{eqnarray*}
\begin{aligned}
P\text{=}\sum_{i=1}^I P_i\text{=}\sum_{i=1}^{I}\frac{|\widehat{\mathcal{P}}_i^{+}\cap \mathcal{P}_i^+|}{|\widehat{\mathcal{P}}_i^{+}|}\times 100\%,~
R\text{=}\sum_{i=1}^I R_i\text{=}\sum_{i=1}^{I}\frac{|\widehat{\mathcal{P}}_i^{+}\cap \mathcal{P}_i^+|}{|\mathcal{P}_i^+|}\times 100\%,~
F1\text{=}\sum_{i=1}^I\frac{2P_i R_i}{P_i+R_i}.
\end{aligned}
\end{eqnarray*}

For each method, the embedding model is learned on the training admissions with 10-fold cross validation. 
In particular, the model is learned on 9-fold of training admission, validated on the remaining fold, and tested on the testing admission. 
The averaged top-$L$ measurements of each model, with $L=1,~3,~5,~10$, are recorded in Table~\ref{tab:cmp}.\footnote{The 90\%-confidence intervals of all the measurements are calculated as well. 
However, we find that the range of the confidence interval for all the methods and measurements are smaller than $\pm 0.15\%$, which means that our method and the alternatives have stable performance on our datasets.
Therefore, here we just show averaged measurements in Table~\ref{tab:cmp}.}
We find that the proposed E+SA+MA method outperforms the alternatives consistently across different datasets and for all measurements.
Specifically, among existing methods, only BPR is comparable to that of E+Pooling (the simplest variant of our method). 
Considering self-attention or mutual-attention mechanism indeed boosts the recommendation result in most situations, as shown in the rows of E+SA and E+MA. 
Accordingly, combining these two mechanisms jointly ($i.e.$, our E+SA+MA) can achieve the best performance.

\subsection{Robustness to hyperparameters}\label{ssec:robust}
\begin{figure}[t]
    \centering
    \subfigure[F1 v.s. dimension $M$, with $\alpha=0.1$, $K=8$]{
    \includegraphics[height=4.7cm]{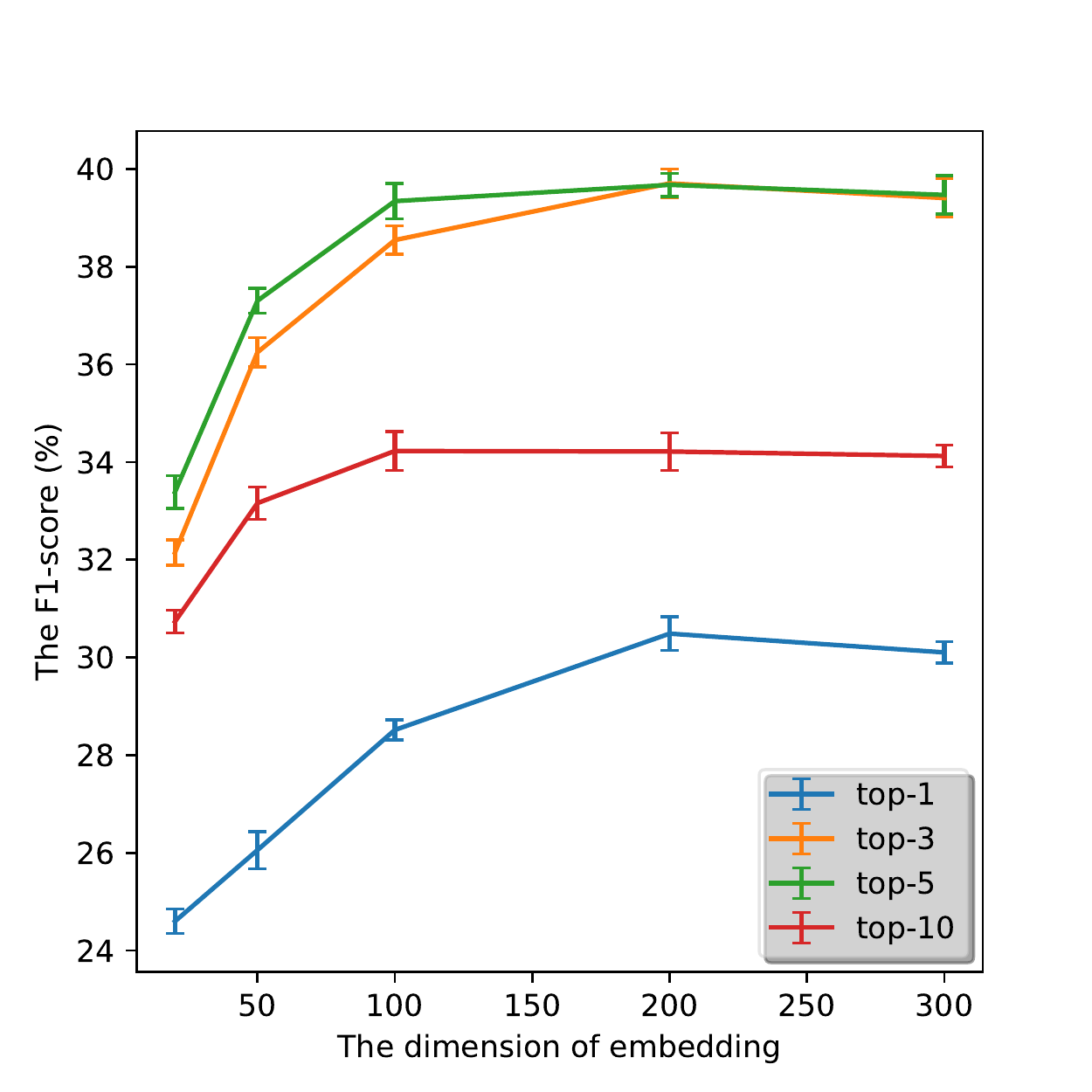}\label{fig:robust_M}
    }
    \subfigure[F1 v.s. weight $\alpha$, with $M=200$, $K=8$]{
    \includegraphics[height=4.7cm]{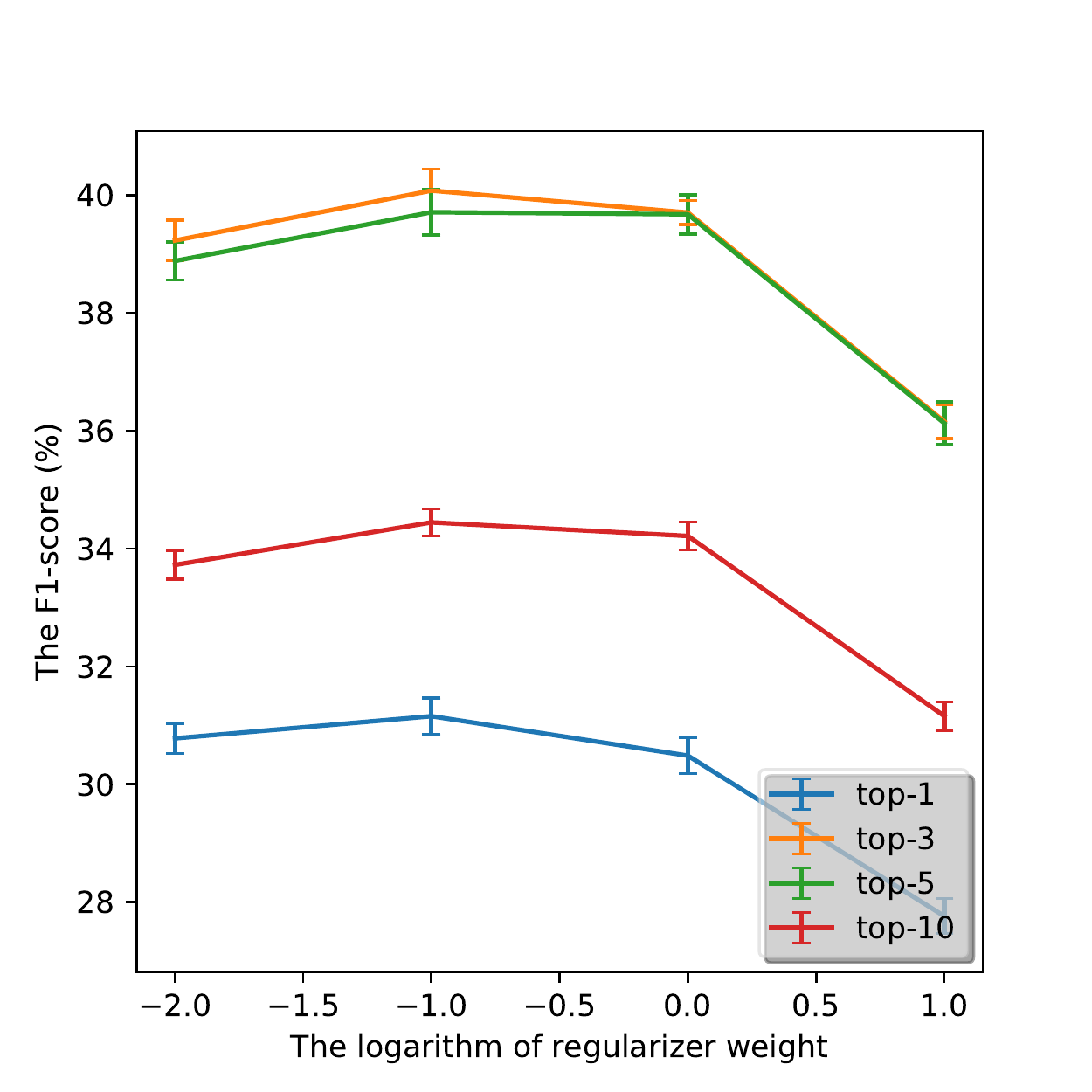}\label{fig:robust_alpha}
    }
    \subfigure[F1 v.s. head number $K$, with $M=200$, $\alpha=0.1$]{
    \includegraphics[height=4.7cm]{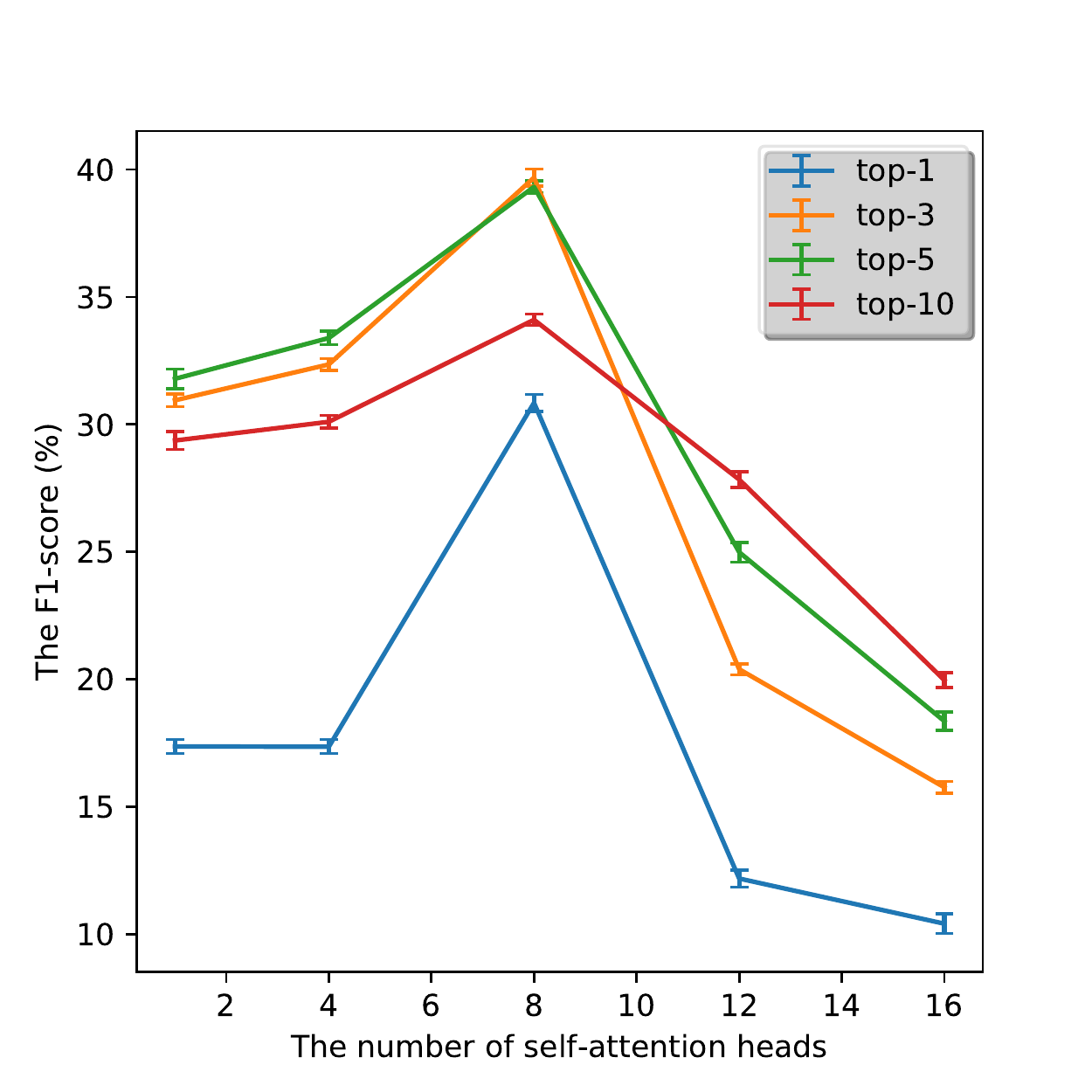}\label{fig:robust_K}
    }
    \caption{Illustrations of the robustness of our method to its hyperparameters, including the dimension of embedding $M$, the weight of the mutual-attention regularizer $\alpha$ and the number of heads in the self-attention network $K$.}
    \label{fig:robustness}
\end{figure}

As discussed above, the dimension $M$ of the embedding vectors, the weight of mutual-attention regularizer $\alpha$, and the number of self-attention heads $K$ are important for our method, having significant influence on the recommendation results. 
Figure~\ref{fig:robustness} illustrates the influence of these hyperparameters on the F1-scores derived by our method on the small dataset.
In Figure~\ref{fig:robust_M}, we find that when the dimension of embedding falls in the range from $100$ to $300$, the performance of our method is relatively stable.
Setting $M=200$ achieves slightly better performance.
When the dimension is too small, $e.g.$, $M=20$ or $50$, the embeddings are not representative enough, which leads to under-fitting. 
In Figure~\ref{fig:robust_alpha} we set $\alpha=0.01, 0.1, 1, 10$. 
The proposed method obtains comparable F1-scores when $\alpha$ falls in the range $0.1$ to $1$. 
When $\alpha$ is too large, $e.g.$, $\alpha=10$, the proposed regularizer is too strong and becomes dominant in the loss function. 
As a result, the model suffers from serious model misspecification and the recommendation results degrade accordingly. 
In Figure~\ref{fig:robust_K}, we set $K=1, 4, 8, 12, 16$.
When $K=1$, the self-attention network only contains one attention layer with one head. 
With the increase of $K$, the number of parameters in the self-attention network becomes larger and the model becomes more representative.
When $K=8$, the best performance is achieved.
If we further increase $K$, our model will have too many parameters and suffer from over-fitting.  
The illustrations on the medium and the large datasets are similar to Figure~\ref{fig:robustness}.
Based on the analysis above, we empirically set $M=200$, $\alpha=0.1$ and $K=8$.

\subsection{Interpretability}
\begin{figure}[t]
    \centering
    \subfigure[Descriptions of ICD codes]{
    \tiny{
    \begin{tabular}{l|l}
	\hline\hline
	d5856 & End stage renal disease\\
    d4275 & Cardiac arrest\\
    d51881 & Acute respiratory failure\\
    d4254 & Other primary cardiomyopathies\\
    d42732 & Atrial flutter\\
    d25000 & Diabetes mellitus without complication, type II\\
    d53081 & Esophageal reflux\\
    d2767 & Hyperpotassemia\\
    d42731 & Atrial fibrillation\\
    d2724 & Other and unspecified hyperlipidemia\\
    d3659 & Unspecified glaucoma\\
    \hline
    p3893 & Venous catheterization, not elsewhere classified\\
    p9960 & Cardiopulmonary resuscitation, not otherwise specified\\
    p9671 & Continuous invasive mechanical ventilation\\
    p9604 & Insertion of endotracheal tube\\
    p3995 & Hemodialysis\\
	\hline\hline
	\end{tabular}
    }
    }
    \subfigure[$\bm{\mu}_{\mathcal{D}_i}$]{
    \includegraphics[height=5cm]{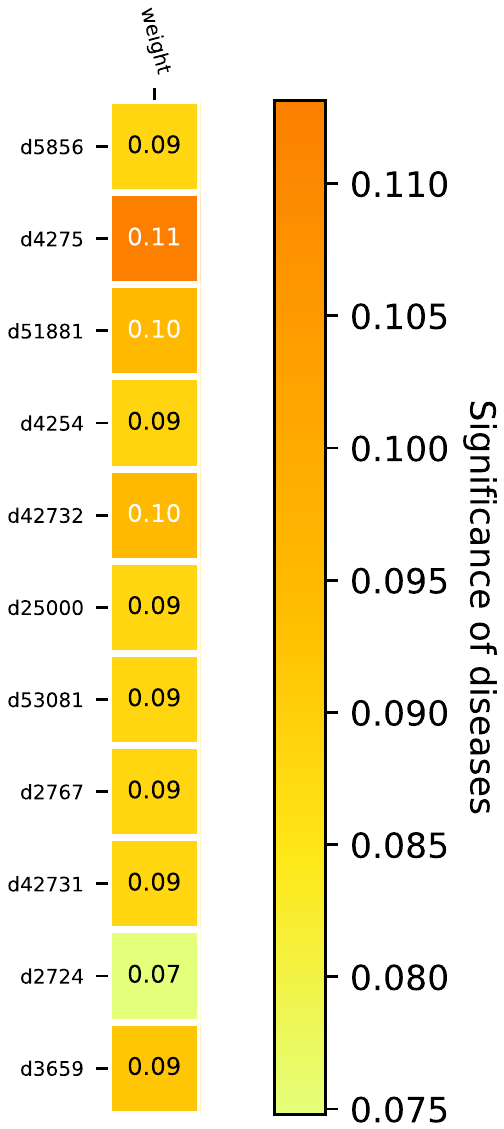}\label{fig:b}
    }
    \subfigure[$\widehat{\bm{T}}_i$]{
    \includegraphics[height=5cm]{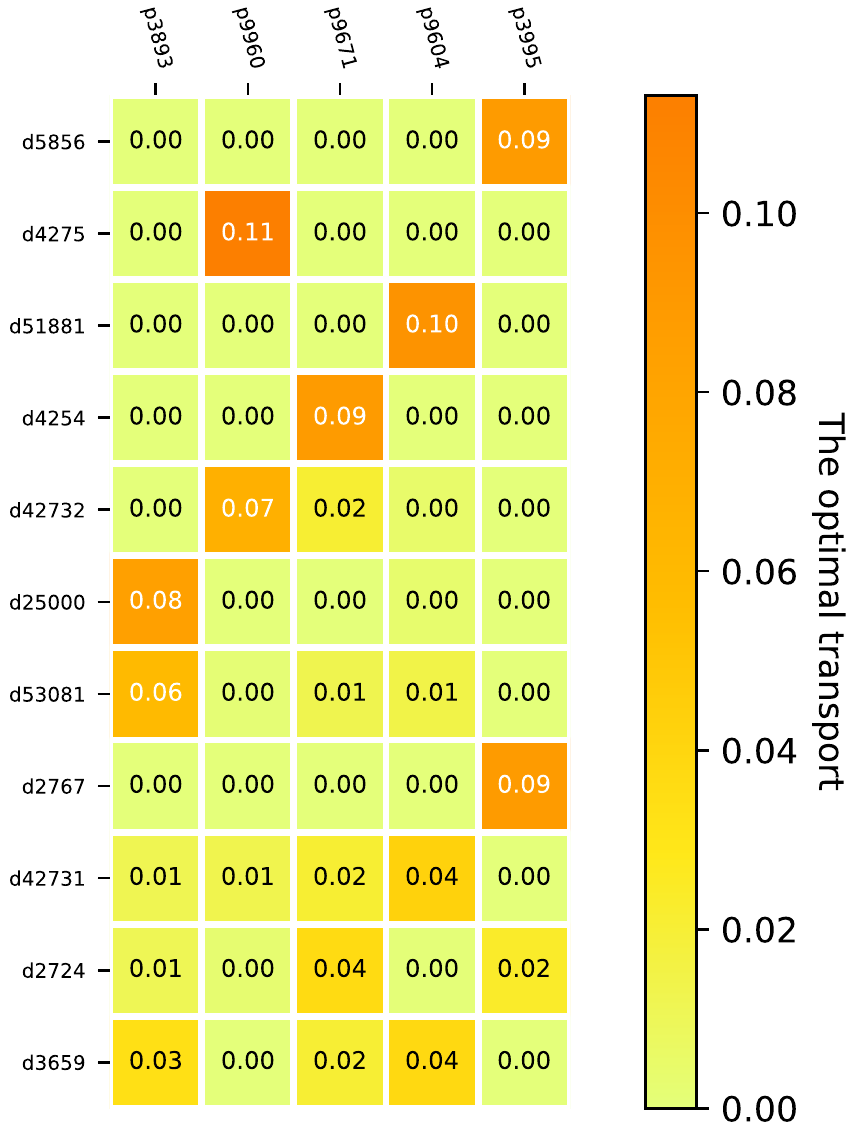}\label{fig:c}
    }\\
    \subfigure[Descriptions of ICD codes]{
    \tiny{
    \begin{tabular}{l|l}
	\hline\hline
	d0389 & Unspecified septicemia\\
    d78552 & Septic shock\\
    d43411 & Cerebral embolism with cerebral infarction\\
    d51884 & Acute and chronic respiratory failure\\
    d41071 & Subendocardial infarction, initial episode of care\\
    d5845 & Acute kidney failure with lesion of tubular necrosis\\
    d4280 & Congestive heart failure, unspecified\\
    d5990 & Urinary tract infection, site not specified\\
    dV5867 & Long-term (current) use of insulin\\
    d2948 & Other persistent mental disorders\\
    d4019 & Unspecified essential hypertension\\
    dV4501 & Cardiac pacemaker\\
    d99592 & Severe sepsis\\ \hline
    p17 & Infusion of vasopressor agent\\
    p9604 & Insertion of endotracheal tube\\
    p9672 & Continuous invasive mechanical ventilation\\
    p966 & Enteral infusion of concentrated nutritional substances\\
    p3893 & Venous catheterization, not elsewhere classified\\
	\hline\hline
	\end{tabular}
    }
    }
    \subfigure[$\bm{\mu}_{\mathcal{D}_i}$]{
    \includegraphics[height=5cm]{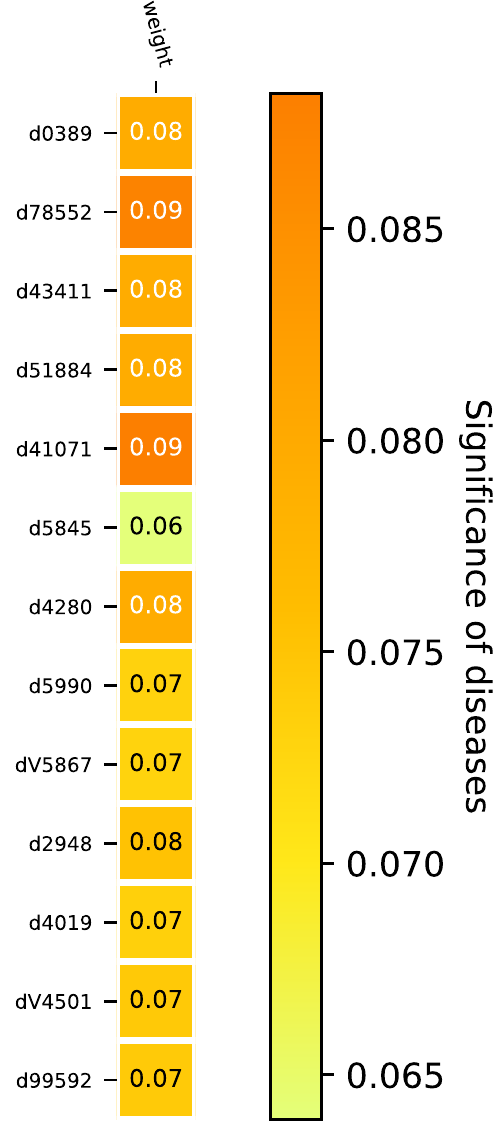}\label{fig:e}
    }
    \subfigure[$\widehat{\bm{T}}_i$]{
    \includegraphics[height=5cm]{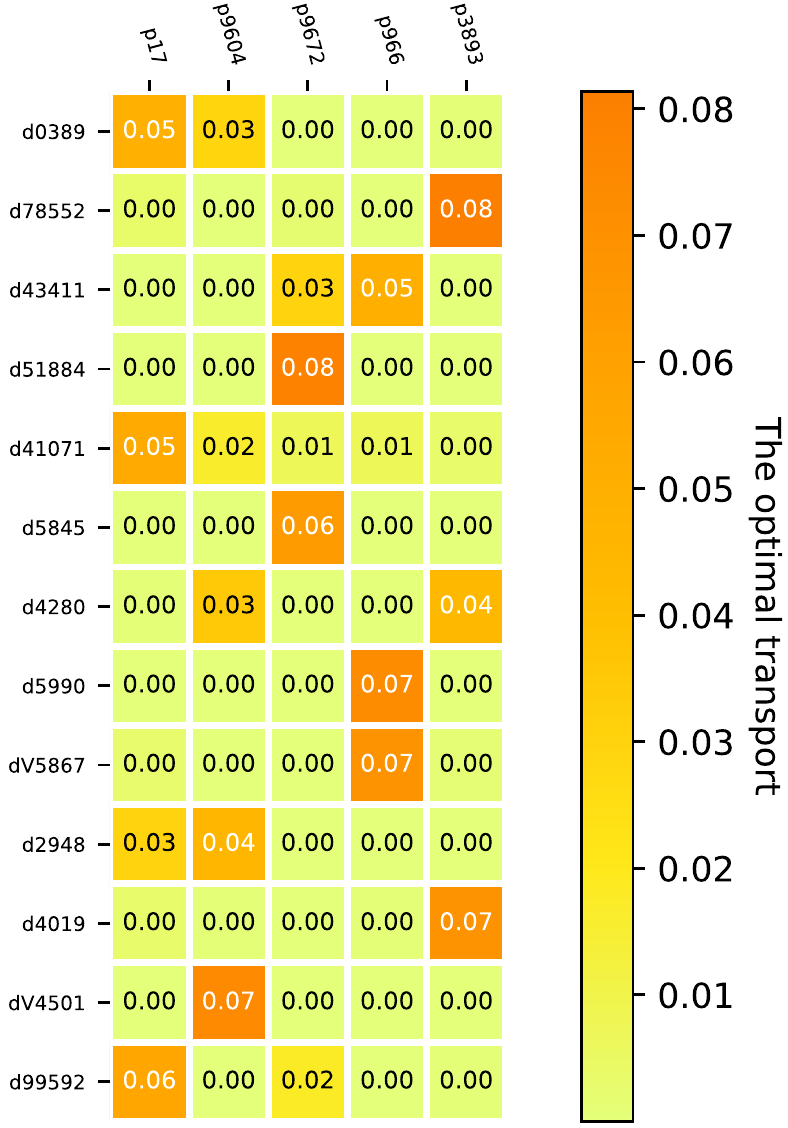}\label{fig:f}
    }\\
    \subfigure[Descriptions of ICD codes]{
    \tiny{
    \begin{tabular}{l|l}
	\hline\hline
	d51882 & Other pulmonary insufficiency, not elsewhere classified\\
    d32723 & Obstructive sleep apnea (adult)(pediatric)\\
    d49320 & Chronic obstructive asthma, unspecified\\
    d42830 & Diastolic heart failure, unspecified\\
    d4280 & Congestive heart failure, unspecified\\
    d5849 & Acute kidney failure, unspecified\\
    d2762 & Acidosis\\
    d2760 & Hyperosmolality and/or hypernatremia\\
    d4168 & Other chronic pulmonary heart diseases\\
    \hline
    p3891 & Arterial catheterization\\
    p9671 & Continuous invasive mechanical ventilation\\
    p9604 & Insertion of endotracheal tube\\
    p3893 & Venous catheterization, not elsewhere classified\\
    p9390 & Non-invasive mechanical ventilation\\
	\hline\hline
	\end{tabular}
    }
    }
    \subfigure[$\bm{\mu}_{\mathcal{D}_i}$]{
    \includegraphics[height=4.5cm]{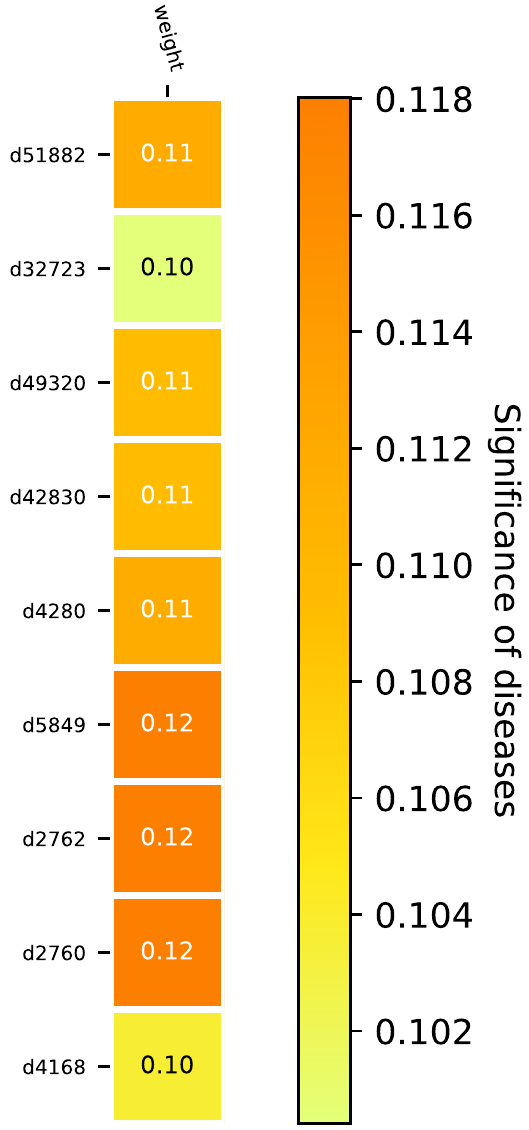}\label{fig:h}
    }
    \subfigure[$\widehat{\bm{T}}_i$]{
    \includegraphics[height=4.5cm]{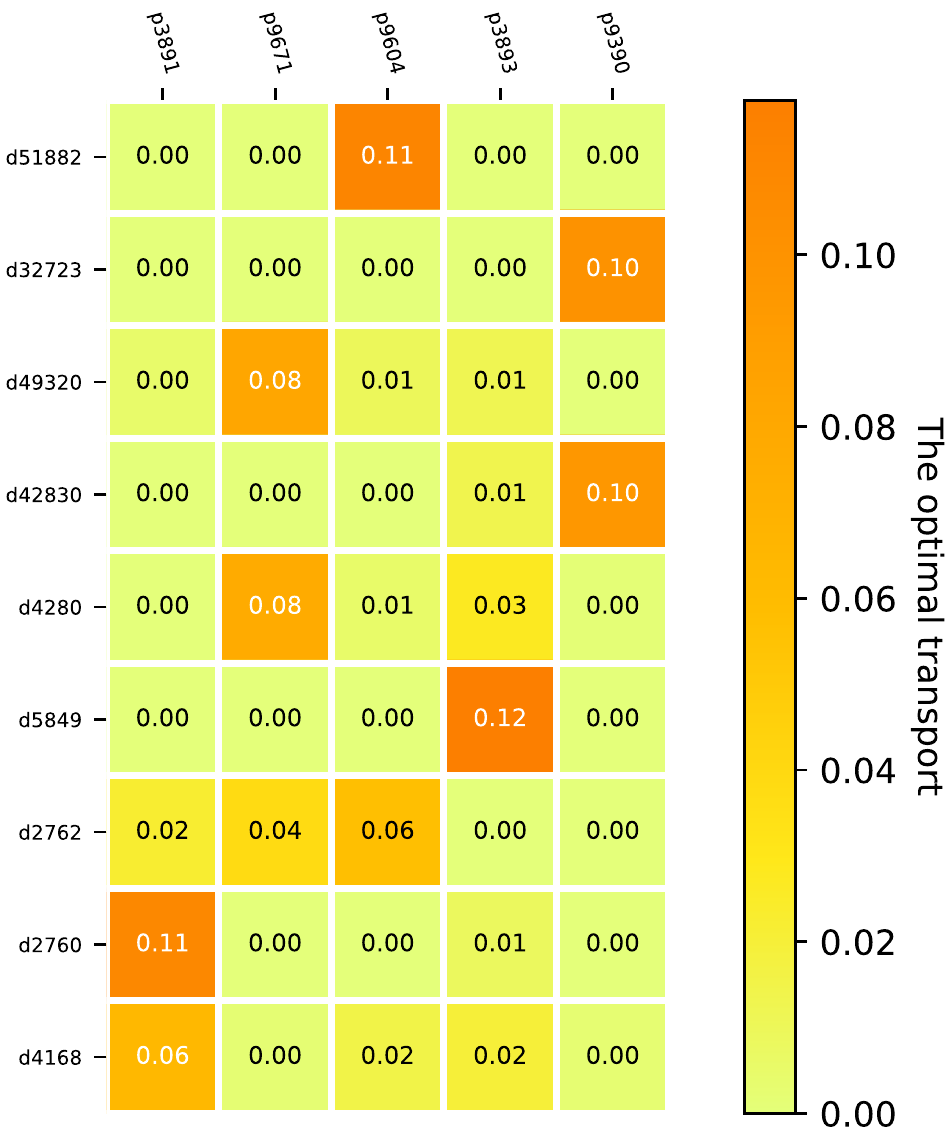}\label{fig:i}
    }
    \caption{Typical examples showing the interpretability of our method. 
    For each row, the descriptions of diagnosed diseases and recommended procedures, the estimated significance of the diseases, and the optimal transport between the diseases and the procedures are shown from left to right.}
    \label{fig:example}
\end{figure}

The embeddings we learn are found to have good interpretability.
Specifically, given a set of diseases, our self-attention network estimates the significance of the diseases. 
Given predicted procedures, we can further calculate the optimal transport between the observed diseases and the predicted procedures explicitly via the mutual-attention regularizer. 
Figure~\ref{fig:example} shows typical procedure recommendation results. 
Each row corresponds to the result of an admission. 
For each admission, we observe some diagnosed diseases and recommended top-5 procedures. 
Given the diagnosed diseases, their significance ($i.e.$, $\bm{\mu}_{\mathcal{D}_i}$) and their optimal transport ($i.e.$,  {$\widehat{\bm{T}}_i$}) to the recommended procedures are shown. 
We find that the significance reflects the seriousness of diseases: in Figure~\ref{fig:b}, the urgent disease ``Cardiac arrest'' is assigned with the highest significance; in Figure~\ref{fig:h}, dangerous diseases like ``Acute kidney failure'' are assigned with high significance. 
Additionally, the optimal transport estimates reasonable clinical relationships between diseases and procedures.
For example, in Figure~\ref{fig:c}, we find that the disease ``End stage renal disease'' will transport to its regular procedure ``Hemodialysis'', and the disease ``Cardiac arrest'' will transport to the suitable procedure ``Cardiopulmonary resuscitation''.
In Figure~\ref{fig:f}, the disease ``Acute and chronic respiratory failure'' may yield procedure ``Continuous invasive mechanical ventilation''.
In Figure~\ref{fig:i}, the disease ``Other pulmonary insufficiency'' may lead to ``Insertion of endotracheal tube''. 
More examples can be found in Appendix~\ref{ap:2}.

\section{Conclusions and Future Work} 
A novel embedding method has been proposed for ICD codes, which has self- and mutual-attention mechanisms and outperforms existing embedding methods in procedure recommendation.
The self-attention network in the proposed method achieves an adaptive fusion strategy of disease embeddings, which estimates the significance of various diseases in different admissions. 
An optimal transport-based mutual-attention regularizer is considered in the training phase of our model, estimating the clinical relationship between the diseases and the procedures appearing in the same admissions. 
These two mechanisms enhance the interpretability of the learned embeddings and improve the recommendation accuracy.
In the future, we plan to design more-effective attention mechanisms and further improve the representation power of embeddings.
Additionally, beyond the MIMIC-III, we will explore the performance of our method in real-world large-scale data.


\bibliography{recsys_icd}

\appendix
\section{Proximal Gradient Method for Optimal Transport}\label{ap:1}
Mathematically, the optimal-transport distance can be defined as follows~\citep{villani2008optimal}:
\begin{definition}
Let $(\Omega, D)$ be an arbitrary space with metric $D$ and $P(\Omega)$ the set of Borel probability measures on $\Omega$. 
For probability measures $u$ and $v$ in $P(\Omega)$, their optimal-transport distance is
\begin{eqnarray}\label{eq:otd}
\begin{aligned}
W(u,v):=\sideset{}{_{\pi\in\Pi(u,v)}}\inf\int_{\Omega\times\Omega}D(x, y)d\pi(x,y),
\end{aligned}
\end{eqnarray}
where $\Pi(u,v)$ is the set of all probability measures on $\Omega\times\Omega$ with $u$ and $v$ as marginals.
\end{definition}
When the metric $D$ is Euclidean, the optimal transport distace corresponds to well-known Wasserstein distance~\citep{villani2008optimal,cuturi2013sinkhorn}.
When $D$ is not a valid metric, the optimal-transport distance corresponds to the classcial Monge–Kantorovich transportation problem.

For the discrete case in our work, given a set of diseases $\mathcal{D}_i$, a set of procedures $\mathcal{P}_i^+$, and their distributions $\bm{\mu}_{\mathcal{D}_i}\in\Sigma^{|\mathcal{D}_i|}$ and $\bm{\mu}_{\mathcal{P}_i^+}\in\Sigma^{|\mathcal{P}_i^+|}$, the definition in (\ref{eq:otd}) can be reformulated as (\ref{eq:loss_ot}), where the cost matrix $\bm{C}(\mathcal{D}_i,\mathcal{P}_i^+)$ is calculated via (\ref{eq:cost}).

\begin{algorithm}[t]
	\caption{Proximal gradient method for optimal transport}
	\label{alg2}
	\begin{algorithmic}[1]
		\STATE \textbf{Input:} Cost matrix $\bm{C}_{\theta}(\mathcal{D}_i,\mathcal{P}_i^+)$, the distributions of diseases and procedures $\bm{\mu}_{\mathcal{D}_i}$ and $\bm{\mu}_{\mathcal{P}_i^+}$, the weight of proximal term $\beta=0.5$.
        \STATE \textbf{Output:} Optimal transport $\bm{T}$
        \STATE Calculate $\bm{G} = \exp(-\frac{\bm{C}_{\theta}(\mathcal{D}_i,\mathcal{P}_i^+)}{\beta})$, and initialize $\bm{a}=\bm{1}_{|\mathcal{D}_i|}$, $\bm{T} = \bm{\mu}_{\mathcal{D}_i}\bm{\mu}_{\mathcal{P}_i^+}^{\top}$
        \WHILE{not converge}
        \STATE $\bm{K} = \bm{G}\odot \bm{T}$ 
        \WHILE{not converge}
        \STATE \CommentSty{Sinkhorn-Knopp Iteration:} $\bm{b}=\frac{\bm{\mu}_{\mathcal{P}_i^+}}{\bm{K}^{\top}\bm{a}}$, and then $\bm{a}=\frac{\bm{\mu}_{\mathcal{D}_i}}{\bm{K}\bm{b}}$.
        \ENDWHILE
        \STATE $\bm{T}=(\bm{a}\bm{b}^{\top})\odot \bm{K}$
        \ENDWHILE
	\end{algorithmic}
\end{algorithm}

We solve (\ref{eq:loss_ot}) via the proximal gradient method proposed in~\citep{xie2018fast}. 
In particular, this method solves (\ref{eq:loss_ot}) iteratively. 
In the $j$-th iteration, a proximal term is added to original problem as
\begin{eqnarray}\label{eq:prox}
\begin{aligned}
\bm{T}^{(j)}=\arg\sideset{}{_{\bm{T}\in \Pi(\bm{\mu}_{\mathcal{D}_i},\bm{\mu}_{\mathcal{P}_i^{+}})}}\min\langle \bm{C}_{\theta}(\mathcal{D}_i, \mathcal{P}_i^{+}), \bm{T}\rangle + \beta\mbox{KL}(\bm{T}\lVert\bm{T}^{(j-1)}),
\end{aligned}
\end{eqnarray}
where $\bm{T}^{(j-1)}$ is the optimal transport learned in previous iteration, and the proximal term is $\mbox{KL}(\bm{T}\lVert\bm{T}^{(j-1)})=\sum_{d,p}t_{dp}\log\frac{t_{dp}}{t_{dp}^{(j-1)}}$. 

The optimization problem in (\ref{eq:prox}) can be rewritten as
\begin{eqnarray}\label{eq:sinkhorn}
\begin{aligned}
\bm{T}^{(j)}=\arg\sideset{}{_{\bm{T}\in \Pi(\bm{\mu}_{\mathcal{D}_i},\bm{\mu}_{\mathcal{P}_i^{+}})}}\min\langle \bm{C}_{\theta}(\mathcal{D}_i, \mathcal{P}_i^{+}) - \beta\log \bm{T}^{(j-1)}, \bm{T}\rangle + \beta\langle\log\bm{T}, \bm{T}\rangle,
\end{aligned}
\end{eqnarray}
where the second term is the entropy regularizer used in~\citep{cuturi2013sinkhorn}. 
Accordingly, Sinkhorn-Knopp iteration can be applied to solve (\ref{eq:sinkhorn}) effectively.
In summary, the proposed proximal gradient method is shown in Algorithm~\ref{alg2}.
This method has linear convergence and has good numerical stability.
More detailed analysis can be found in~\citep{xie2018fast,xu2019gromov}.


\section{Typical Examples of Learning Results}\label{ap:2}
\begin{figure}[t]
    \centering
    \subfigure[Descriptions of ICD codes]{
    \tiny{
    \begin{tabular}{l|l}
	\hline\hline
	d0543 & Herpetic meningoencephalitis \\
    d51881 & Acute respiratory failure \\
    d3485 & Cerebral edema \\
    d4019 & Unspecified essential hypertension \\
    d2720 & Pure hypercholesterolemia \\
    d4240 & Mitral valve disorders \\ \hline
    p331 & Incision of lung \\
    p9671 & Continuous invasive mechanical ventilation \\
    p9604 & Insertion of endotracheal tube \\
    p966 & Enteral infusion of concentrated nutritional substances \\
    p3893 & Venous catheterization, not elsewhere classified \\
	\hline\hline
	\end{tabular}
    }
    }
    \subfigure[$\bm{\mu}_{\mathcal{D}_i}$]{
    \includegraphics[height=3.7cm]{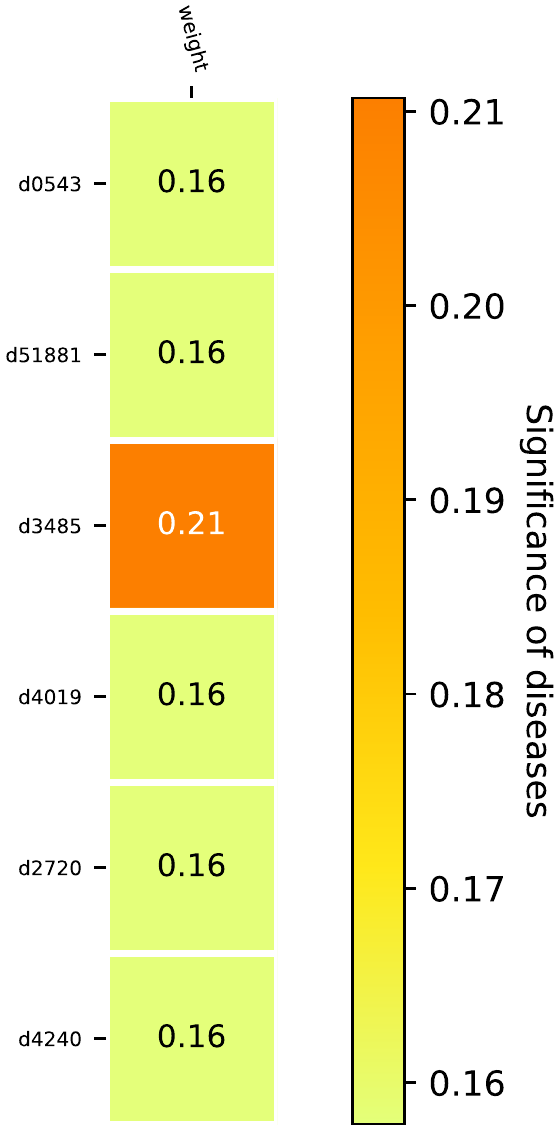}
    }
    \subfigure[$\widehat{\bm{T}}_i$]{
    \includegraphics[height=3.5cm]{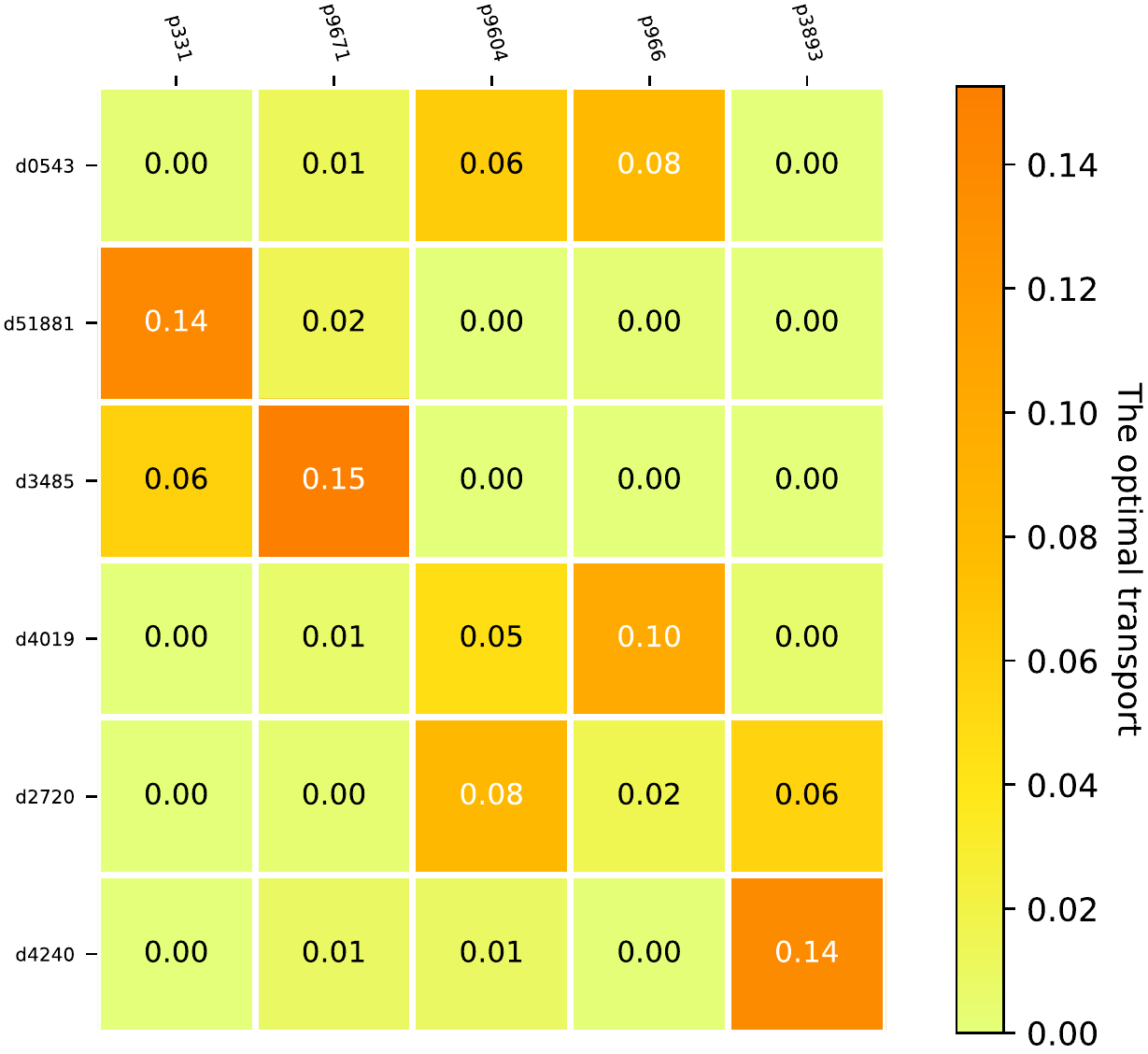}
    }\\
    \subfigure[Descriptions of ICD codes]{
    \tiny{
    \begin{tabular}{l|l}
	\hline\hline
	d2732 & Other paraproteinemias \\
    d5849 & Acute kidney failure, unspecified \\
    d2761 & Hyposmolality and/or hyponatremia \\
    d2767 & Hyperpotassemia \\
    d5839 & Nephritis and nephropathy \\
    d2859 & Anemia, unspecified \\
    d7102 & Sicca syndrome \\ \hline
    p9971 & Therapeutic plasmapheresis \\
    p3893 & Venous catheterization, not elsewhere classified \\
    p5523 & Closed [percutaneous] [needle] biopsy of kidney \\
    p3895 & Venous catheterization for renal dialysis \\
    p3995 & Hemodialysis \\
	\hline\hline
	\end{tabular}
    }
    }
    \subfigure[$\bm{\mu}_{\mathcal{D}_i}$]{
    \includegraphics[height=4cm]{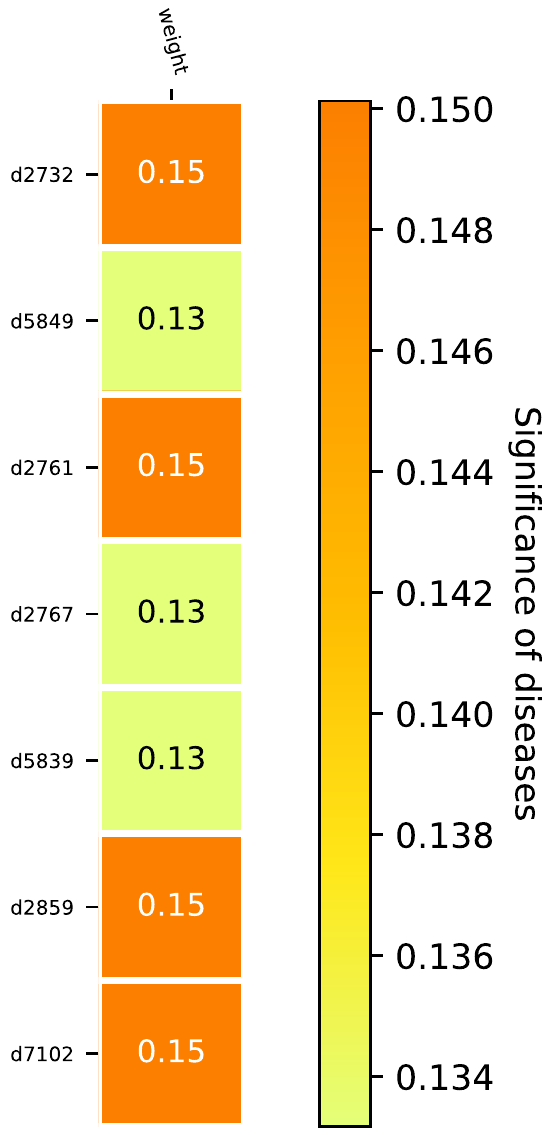}
    }
    \subfigure[$\widehat{\bm{T}}_i$]{
    \includegraphics[height=4cm]{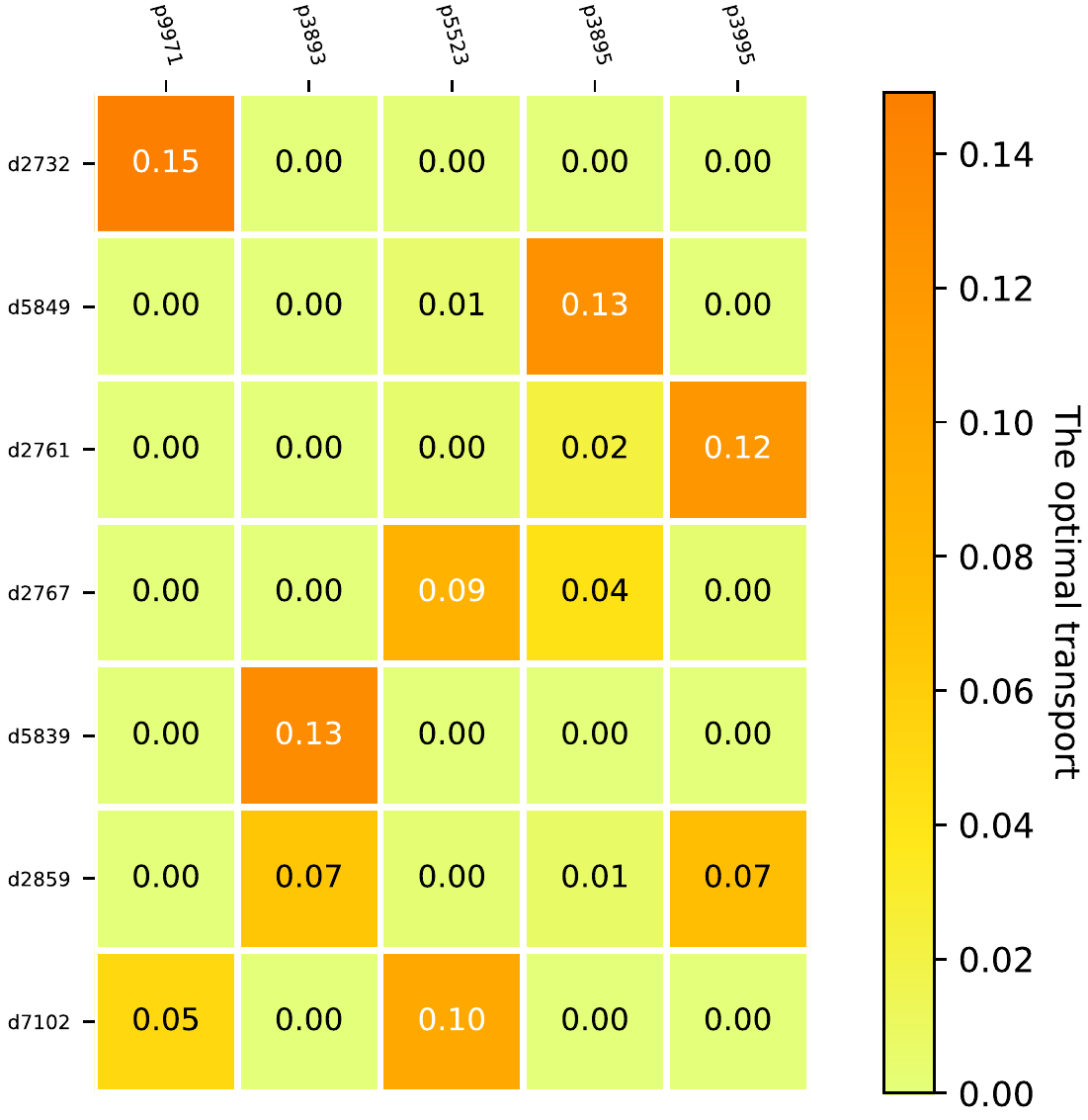}
    }\\
    \subfigure[Descriptions of ICD codes]{
    \tiny{
    \begin{tabular}{l|l}
	\hline\hline
	d4240 & Mitral valve disorders \\
    d5859 & Chronic kidney disease, unspecified \\
    d42731 & Atrial fibrillation \\
    d2449 & Unspecified acquired hypothyroidism \\
    d4019 & Unspecified essential hypertension \\
    d2720 & Pure hypercholesterolemia \\
    dV5861 & Long-term (current) use of anticoagulants \\
    d2859 & Anemia, unspecified \\
    d2749 & Gout, unspecified \\ \hline
    p3733 & Excision or destruction of other lesion or tissue of heart \\
    p8872 & Diagnostic ultrasound of heart \\
    p3523 & Open and replacement of mitral valve with tissue graft \\
    p9904 & Transfusion of packed cells \\
    p3961 & Extracorporeal circulation auxiliary to open heart surgery \\
	\hline\hline
	\end{tabular}
    }
    }
    \subfigure[$\bm{\mu}_{\mathcal{D}_i}$]{
    \includegraphics[height=4cm]{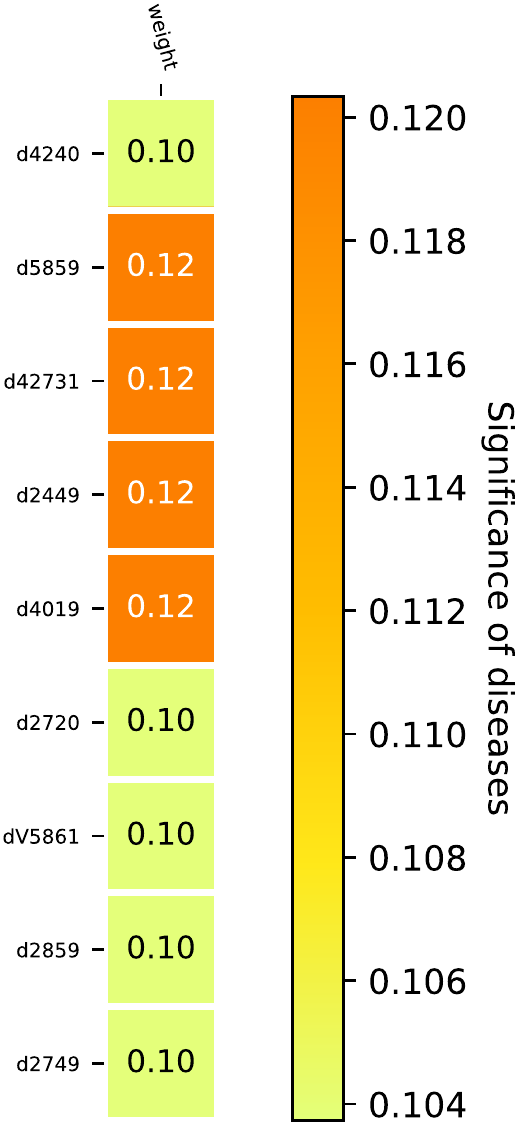}
    }
    \subfigure[$\widehat{\bm{T}}_i$]{
    \includegraphics[height=4cm]{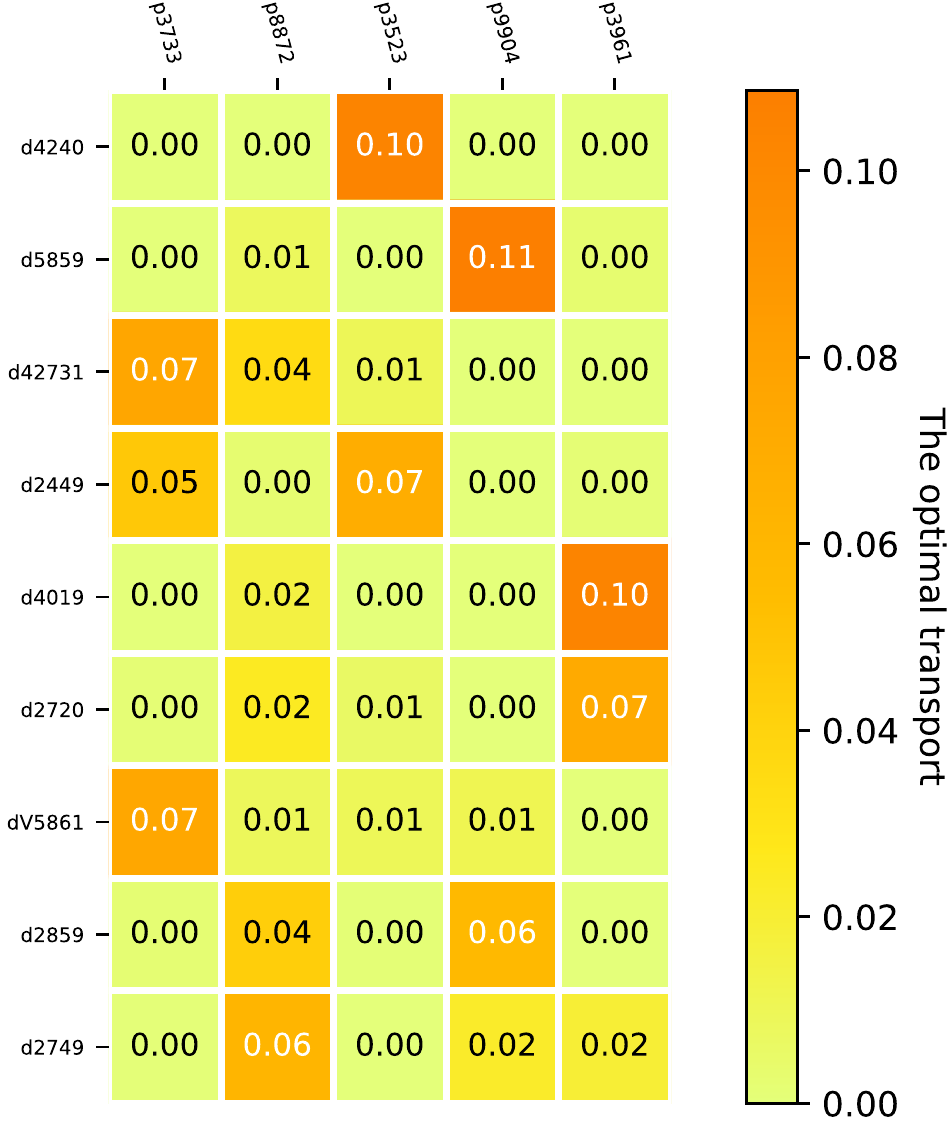}
    }
    \caption{Typical examples showing the interpretability of our method. 
    For each row, the descriptions of diagnosed diseases and recommended procedures, the estimated significance of the diseases, and the optimal transport between the diseases and the procedures are shown from left to right.}
    \label{fig:example2}
\end{figure}

\begin{figure}[t]
    \centering
    \subfigure[Descriptions of ICD codes]{
    \tiny{
    \begin{tabular}{l|l}
	\hline\hline
	d41401 & Coronary atherosclerosis of native coronary artery \\
    d42822 & Chronic systolic heart failure \\
    d4139 & Other and unspecified angina pectoris \\
    d2720 & Pure hypercholesterolemia \\
    d2749 & Gout, unspecified \\
    d4280 & Congestive heart failure, unspecified \\
    d412 & Old myocardial infarction \\ \hline
    p3612 & (Aorto)coronary bypass of two coronary arteries \\
    p8856 & Coronary arteriography using two catheters \\
    p3722 & Left heart cardiac catheterization \\
    p3961 & Circulation auxiliary to open heart surgery \\
    p3615 & Single internal mammary-coronary artery bypass \\
	\hline\hline
	\end{tabular}
    }
    }
    \subfigure[$\bm{\mu}_{\mathcal{D}_i}$]{
    \includegraphics[height=4cm]{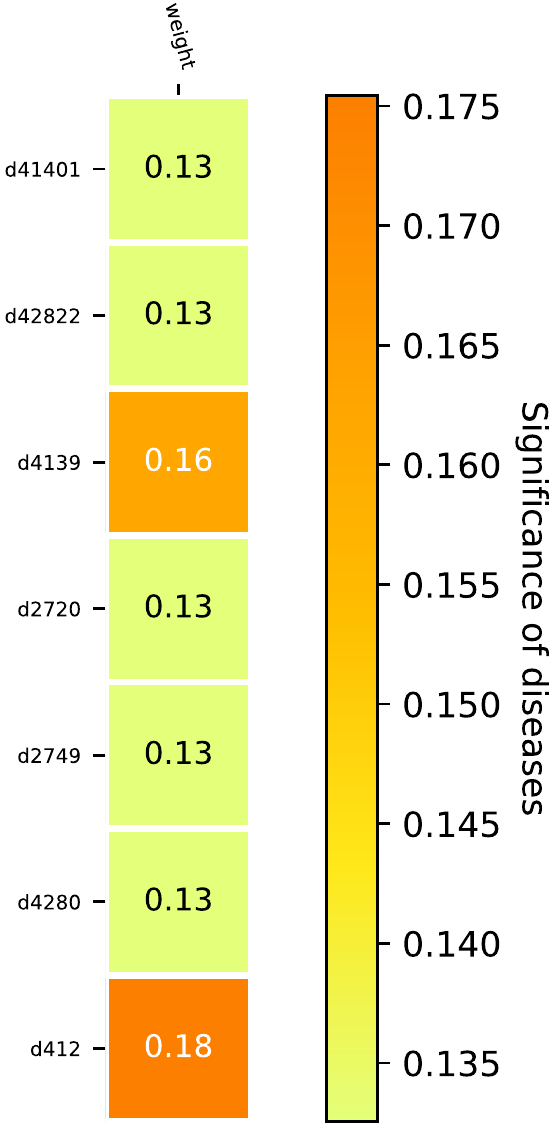}
    }
    \subfigure[$\widehat{\bm{T}}_i$]{
    \includegraphics[height=4cm]{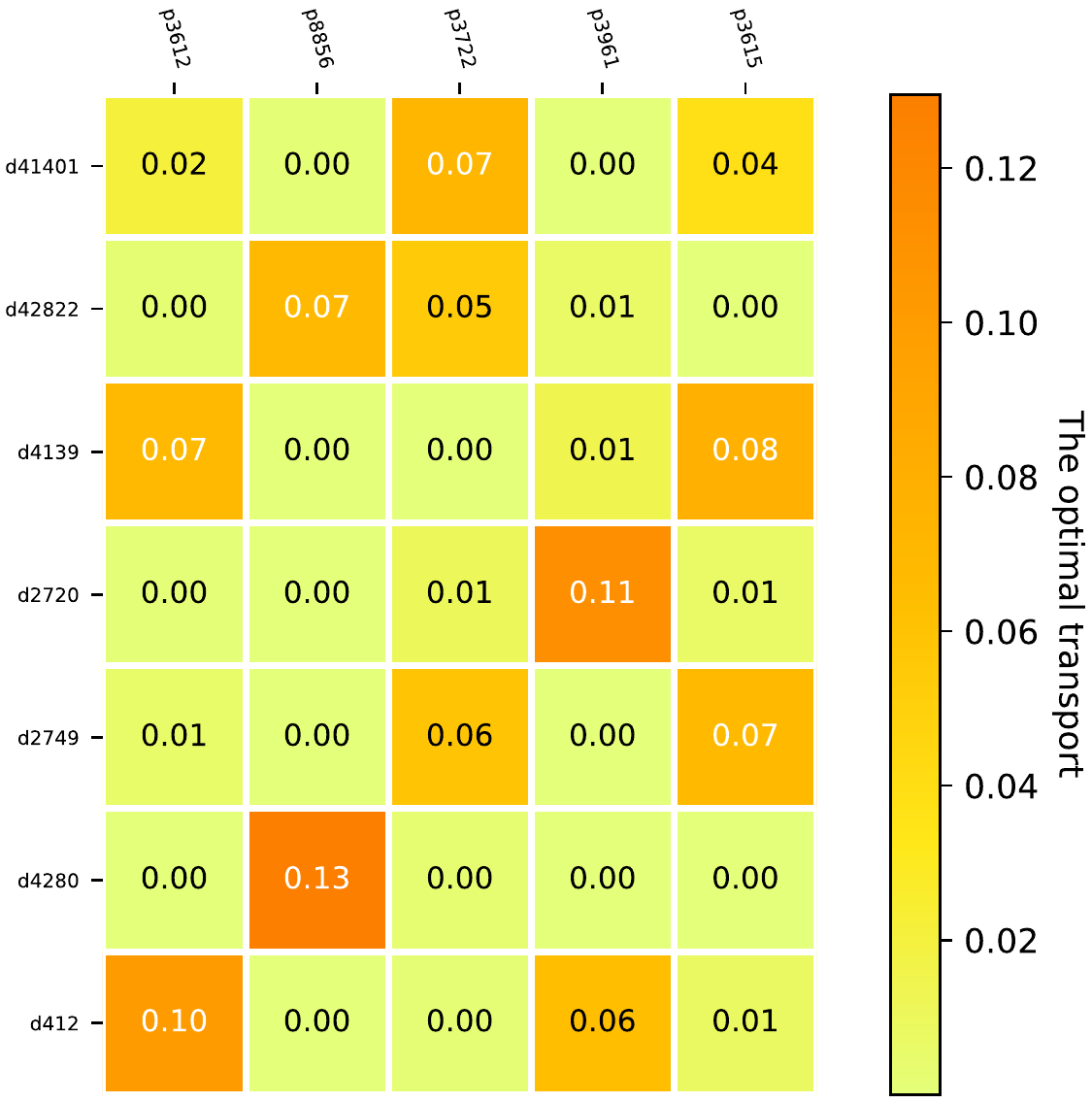}
    }\\
    \subfigure[Descriptions of ICD codes]{
    \tiny{
    \begin{tabular}{l|l}
	\hline\hline
	dV3401 & Other multiple birth (three or more), mates all liveborn \\
    d76516 & Other preterm infants, 1,500-1,749 grams \\
    d7470 & Patent ductus arteriosus \\
    dV053 & Need for prophylactic vaccination against viral hepatitis \\
    d7742 & Neonatal jaundice associated with preterm delivery \\
    d7706 & Transitory tachypnea of newborn \\
    d76525 & 29-30 completed weeks of gestation \\
    d77081 & Primary apnea of newborn \\
    d77981 & Neonatal bradycardia \\
    p9604 & Insertion of endotracheal tube \\
    p9390 & Non-invasive mechanical ventilation \\
    p966 & Enteral infusion of concentrated nutritional substances \\
    p9983 & Other phototherapy \\
    p9955 & Prophylactic administration of vaccine \\
	\hline\hline
	\end{tabular}
    }
    }
    \subfigure[$\bm{\mu}_{\mathcal{D}_i}$]{
    \includegraphics[height=4cm]{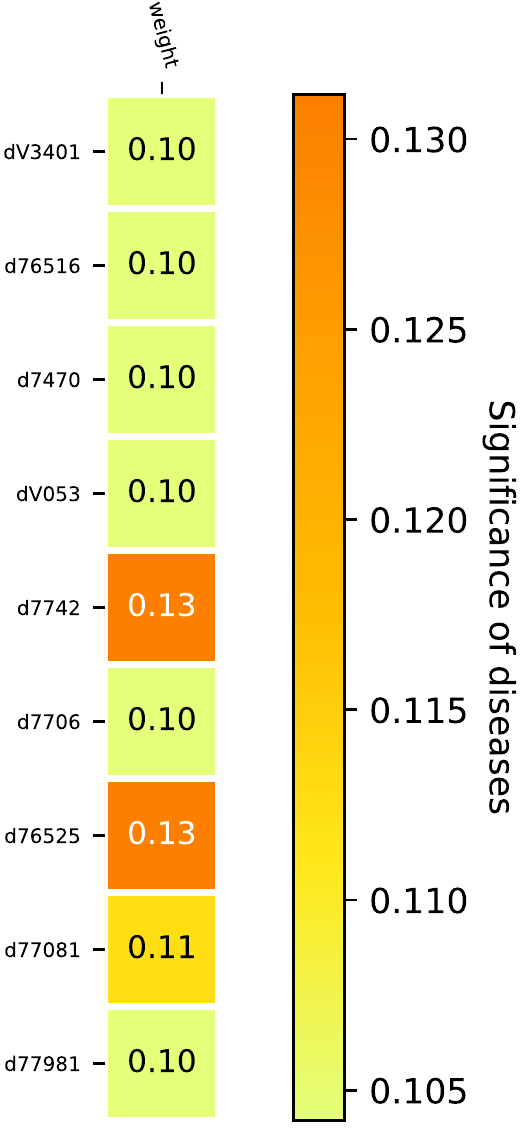}
    }
    \subfigure[$\widehat{\bm{T}}_i$]{
    \includegraphics[height=4cm]{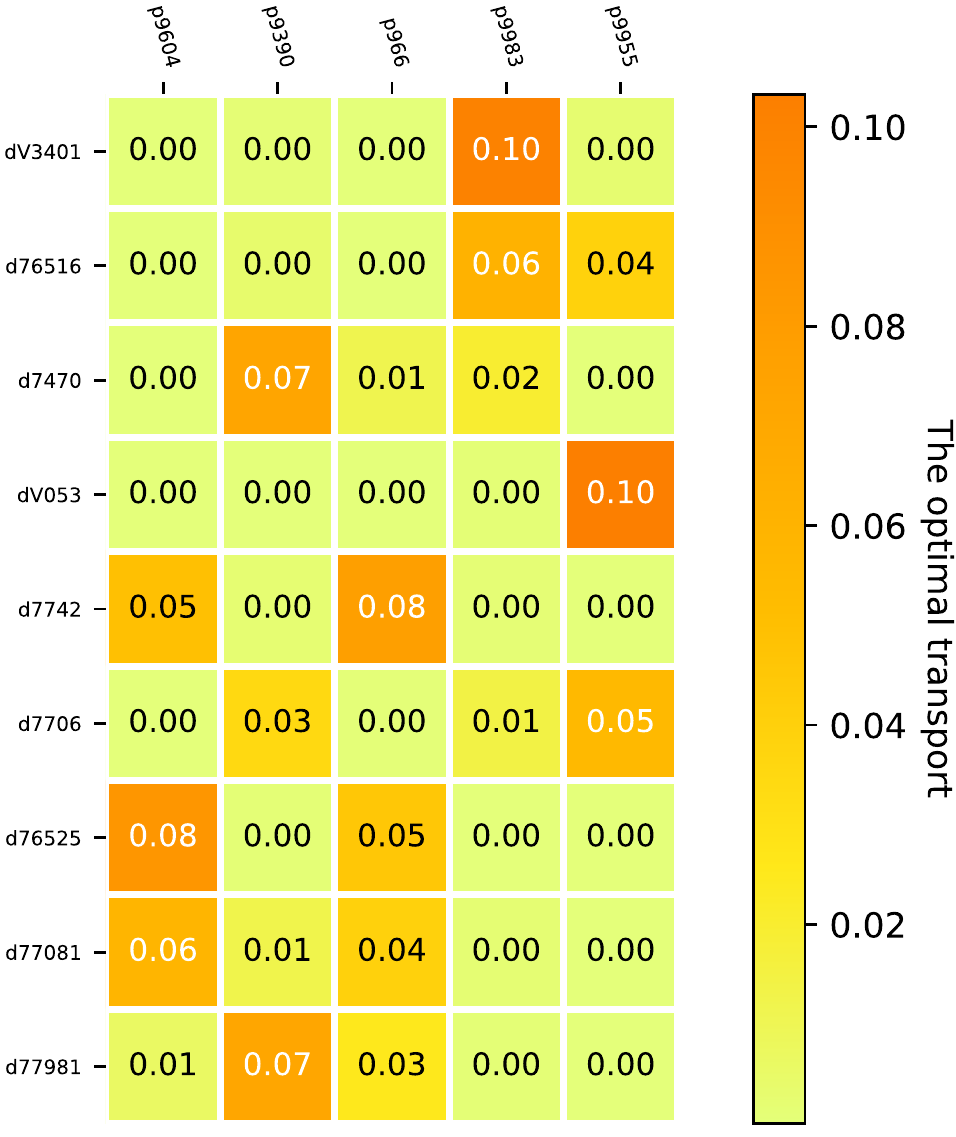}
    }\\
    \subfigure[Descriptions of ICD codes]{
    \tiny{
    \begin{tabular}{l|l}
	\hline\hline
	d41401 & Coronary atherosclerosis of native coronary artery \\
    d4019 & Unspecified essential hypertension \\
    d2720 & Pure hypercholesterolemia \\ \hline
    p3612 & (Aorto)coronary bypass of two coronary arteries \\
    p8856 & Coronary arteriography using two catheters \\
    p3722 & Left heart cardiac catheterization \\
    p3961 & Circulation auxiliary to open heart surgery \\
    p3615 & Single internal mammary-coronary artery bypass \\
	\hline\hline
	\end{tabular}
    }
    }
    \subfigure[$\bm{\mu}_{\mathcal{D}_i}$]{
    \includegraphics[height=3cm]{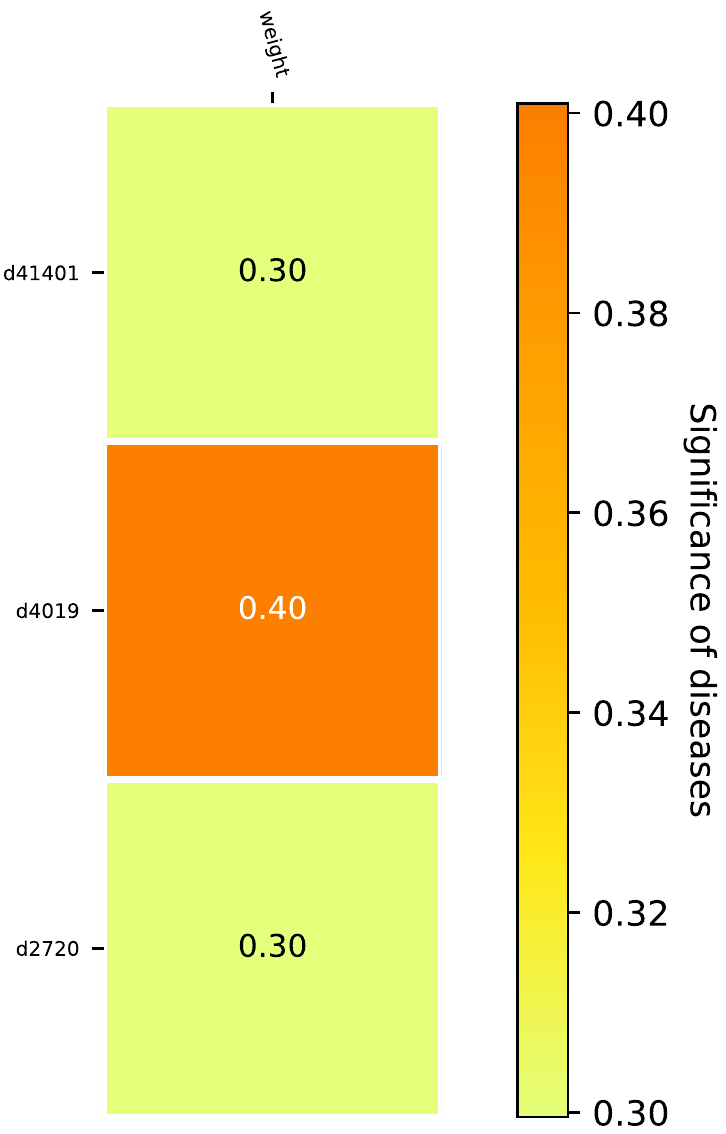}
    }
    \subfigure[$\widehat{\bm{T}}_i$]{
    \includegraphics[height=3cm]{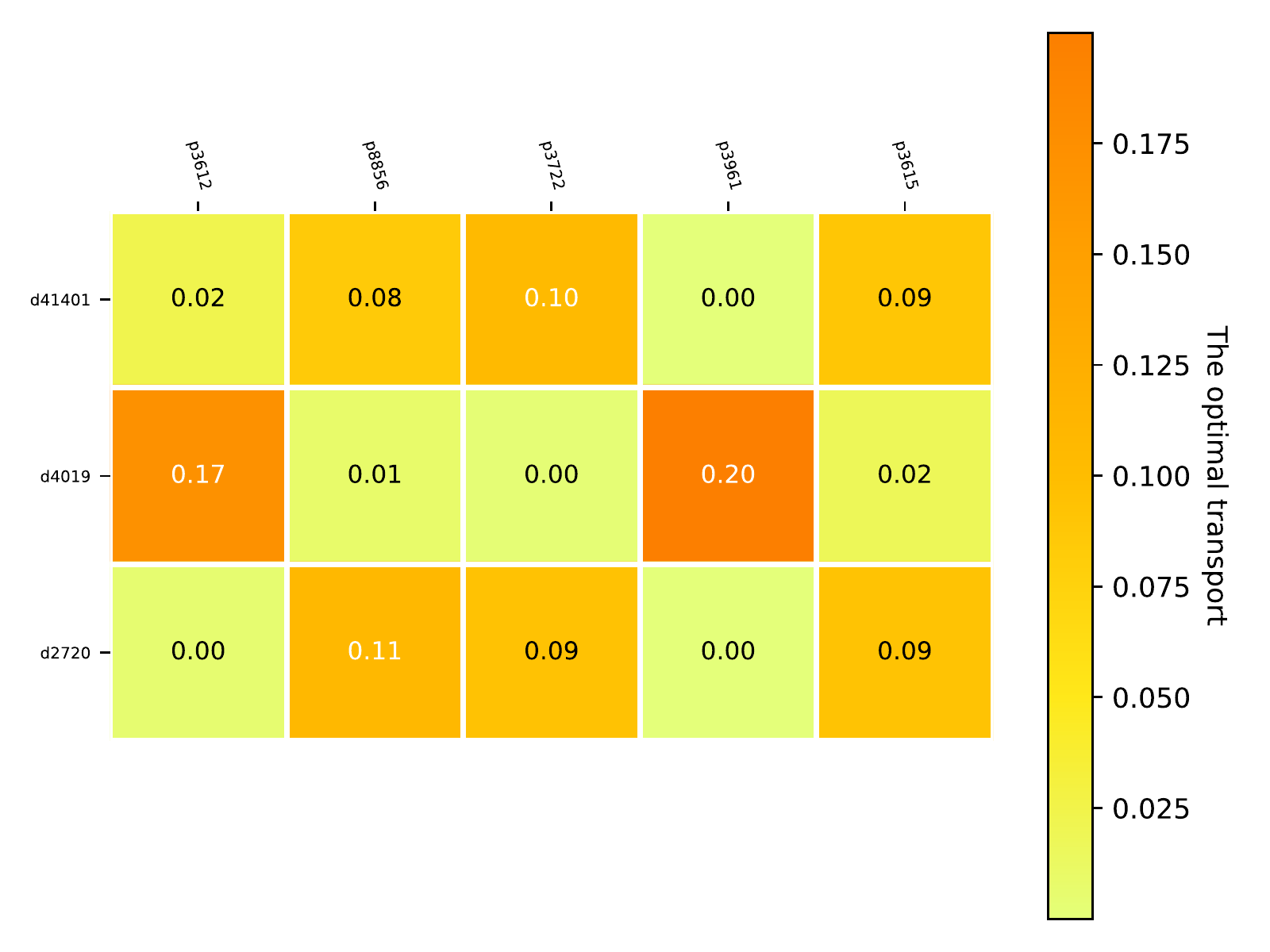}
    }
    \caption{Typical examples showing the interpretability of our method. 
    For each row, the descriptions of diagnosed diseases and recommended procedures, the estimated significance of the diseases, and the optimal transport between the diseases and the procedures are shown from left to right.}
    \label{fig:example3}
\end{figure}

In Figures~\ref{fig:example2} and \ref{fig:example3}, we visualize the learning results of 6 admissions. 
For each admission, the significance of its diagnosed diseases ($i.e.$, $\bm{\mu}_{\mathcal{D}_i}$) and the optimal transport between the diseases and the recommended top-5 procedures ($i.e.$, $\widehat{\bm{T}}_i$) are shown. 
We can find that the significance learned by our method often indicates the main diseases or the most serious diseases in the admissions. 
The nonzero elements in the optimal transports often correspond to the pairs of the diseases and their related procedures. 

\end{document}